\documentclass[sigconf, screen]{acmart}

\AtBeginDocument{%
  \providecommand\BibTeX{{%
    \normalfont B\kern-0.5em{\scshape i\kern-0.25em b}\kern-0.8em\TeX}}}

\usepackage{hyperref}
\usepackage{boldline}
\usepackage{color,colortbl}
\usepackage{bigstrut}
\usepackage[ruled]{algorithm2e}
\usepackage{amsmath,amsfonts}
\usepackage{amsmath}
\usepackage{array}
\usepackage{algorithmic}
\usepackage{balance}
\usepackage{blindtext}
\usepackage{booktabs}
\usepackage{caption}
\usepackage{hyperref}
\usepackage[noabbrev]{cleveref}
\usepackage{comment}
\usepackage{enumitem}
\usepackage{eqparbox}
\usepackage{fancybox}
\usepackage{fancyvrb}
\usepackage{framed}
\usepackage{graphicx}
\usepackage{ifthen}
\usepackage{listings}
\usepackage{mathrsfs}
\usepackage{mdwmath}
\usepackage{multirow}
\usepackage{pifont}
\usepackage{stfloats}
\usepackage{textcomp}
\usepackage{url}
\usepackage{xspace}
\usepackage[normalem]{ulem}
\usepackage[framemethod=TikZ]{mdframed}
\usepackage{caption}
\usepackage{subcaption}
\usepackage{tabularx}
\usepackage{adjustbox}

\usepackage{tcolorbox}
\tcbuselibrary{listings,skins}
\usepackage{mathtools}
\usepackage{blindtext}
\usepackage{lstautogobble}
\usepackage{tikz}
\usepackage{lipsum}

\DeclareGraphicsExtensions{.pdf,.jpeg,.png}

\lstdefinestyle{customc}{
  belowcaptionskip=1\baselineskip,
  xleftmargin=17pt,
  xrightmargin=3pt,
  language=C,
  showstringspaces=false,
  basicstyle=\scriptsize\ttfamily,
  keywordstyle=\bfseries\color{purple!40!black},
  commentstyle=\itshape\color{blue},
  identifierstyle=\color{black},
  stringstyle=\color{orange},
  morekeywords={~!+-*&^/\%},
  numbers=left,
  stepnumber=1,
}

\lstdefinestyle{customcnew}{
  belowcaptionskip=1\baselineskip,
  xleftmargin=17pt,
  xrightmargin=3pt,
  language=C,
  showstringspaces=false,
  basicstyle=\footnotesize\ttfamily,
  keywordstyle=\bfseries\color{purple!40!black},
  commentstyle=\itshape\color{blue},
  identifierstyle=\color{black},
  stringstyle=\color{orange},
  morekeywords={~!+-*&^/\%},
  numbers=left,
  stepnumber=1,
}

\lstdefinestyle{customcwonumberscriptsize}{
  belowcaptionskip=1\baselineskip,
  xleftmargin=0pt,
  language=C,
  showstringspaces=false,
  basicstyle=\scriptsize\ttfamily,
  keywordstyle=\bfseries\color{purple!40!black},
  commentstyle=\itshape\color{blue},
  identifierstyle=\color{black},
  stringstyle=\color{orange},
  morekeywords={~!+-*&^/\%},
  numbers=none,
  stepnumber=2,
}

\lstdefinestyle{customcwonumbersmall}{
  belowcaptionskip=1\baselineskip,
  xleftmargin=0pt,
  language=C,
  showstringspaces=false,
  basicstyle=\small\ttfamily,
  keywordstyle=\bfseries\color{purple!40!black},
  commentstyle=\itshape\color{blue},
  identifierstyle=\color{black},
  stringstyle=\color{orange},
  morekeywords={~!+-*&^/\%},
  numbers=none,
  stepnumber=2,
}

\lstdefinestyle{customcwonumber}{
  belowcaptionskip=1\baselineskip,
  xleftmargin=\parindent,
  language=C,
  showstringspaces=false,
  basicstyle=\scriptsize\ttfamily,
  keywordstyle=\bfseries\color{purple!40!black},
  commentstyle=\itshape\color{blue},
  identifierstyle=\color{black},
  stringstyle=\color{orange},
  morekeywords={~!+-*&^/\%},
  numbers=none,
  stepnumber=2,
}

\lstdefinestyle{customcppwonumber}{
  belowcaptionskip=1\baselineskip,
  xleftmargin=\parindent,
  language=C++,
  showstringspaces=false,
  basicstyle=\Large\ttfamily,
  keywordstyle=\bfseries\color{purple!40!black},
  commentstyle=\itshape\color{blue},
  identifierstyle=\color{black},
  stringstyle=\color{orange},
  morekeywords={~!+-*&^/\%},
  numbers=none,
  stepnumber=2,
}

\newcommand*\circled[1]{\tikz[baseline=(char.base)]{
            \node[shape=circle,draw,inner sep=1pt] (char) {#1};}}

\usetikzlibrary{shadows}          
\newmdenv[
    tikzsetting= {fill=blueish},
    skipabove=0.33em,
    skipbelow=0.33em,
    linewidth=1pt,
    innerleftmargin=4pt,
    innerrightmargin=4pt,
    innertopmargin=2pt,
    innerbottommargin=2pt,
    linecolor=gray95,
    roundcorner=2pt, 
    shadow=true,
    shadowsize=4pt,
    shadowcolor=gray95
]{questionbox}

\newmdenv[
    tikzsetting= {fill=blueish},
    skipabove=0.33em,
    skipbelow=0.33em,
    linewidth=1pt,
    innerleftmargin=4pt,
    innerrightmargin=4pt,
    innertopmargin=2pt,
    innerbottommargin=2pt,
    linecolor=gray95,
    roundcorner=2pt, 
    shadow=true,
    shadowsize=4pt,
    shadowcolor=gray95
]{answerbox}

\newmdenv[
    skipabove=0.33em,
    skipbelow=0.33em,
    innerleftmargin=4pt,
    innerrightmargin=4pt,
    innertopmargin=2pt,
    innerbottommargin=2pt,
]{lessonbox}

\newenvironment{result}
{\begin{answerbox}}
{\end{answerbox}}

\newcommand{\etal}{\hbox{\emph{et al.}}\xspace}
\newcommand{\eg}{\hbox{\emph{e.g.,}}\xspace}
\newcommand{\ie}{\hbox{\emph{i.e.,}}\xspace}
\newcommand{\wrt}{\hbox{\emph{w.r.t.}}\xspace}

\newcommand{\revision}[1]{{#1}}

\newcommand{\tool}{{\sc CONCORD}\xspace}
\newcommand{\toolbf}{{\sc \textbf{CONCORD}}\xspace}

\newcommand{\RS}[2]{%
    \begin{result}
        \textbf{Result-#1:~}{ #2}%
    \end{result}
}

\newcommand\red[1]{\textcolor[rgb]{1.00,0.00,0.00}{#1}}

\newcommand\javagreen[1]{\textcolor[rgb]{0.25,0.5,0.35}{{#1}}}

\definecolor{blueish}{RGB}{250, 250, 255}
\definecolor{greenish}{RGB}{200, 255, 200}
\definecolor{redish}{RGB}{255, 200, 200}
\definecolor{highlight}{RGB}{175, 255, 100}
\definecolor{darkred}{RGB}{139, 0, 0}
\definecolor{gray95}{gray}{0.05}
\definecolor{rowgray}{RGB}{224, 224, 224}

\setcopyright{acmlicensed}
\acmPrice{15.00}
\acmDOI{10.1145/3597926.3598035}
\acmYear{2023}
\copyrightyear{2023}
\acmSubmissionID{issta23main-p26-p}
\acmISBN{979-8-4007-0221-1/23/07}
\acmConference[ISSTA '23]{Proceedings of the 32nd ACM SIGSOFT International Symposium on Software Testing and Analysis}{July 17--21, 2023}{Seattle, WA, United States}
\acmBooktitle{Proceedings of the 32nd ACM SIGSOFT International Symposium on Software Testing and Analysis (ISSTA '23), July 17--21, 2023, Seattle, WA, United States}
\received{2023-02-16}
\received[accepted]{2023-05-03}

\begin{document}

\title{CONCORD: Clone-aware Contrastive Learning for Source Code}

\author{Yangruibo Ding}
\affiliation{%
  \institution{Columbia University}
  \city{New York}
  \state{NY}
  \country{USA}
}
\author{Saikat Chakraborty}
\affiliation{%
    \institution{Microsoft Research}
    \city{Redmond}
  \state{WA}
  \country{USA}
}

\author{Luca Buratti}
\affiliation{%
    \institution{IBM Research}
    \city{Yorktown Heights}
  \state{NY}
  \country{USA}
}

\author{Saurabh Pujar}
\affiliation{%
    \institution{IBM Research}
    \city{Yorktown Heights}
  \state{NY}
  \country{USA}
}

\author{Alessandro Morari}
\affiliation{%
    \institution{IBM Research}
    \city{Yorktown Heights}
  \state{NY}
  \country{USA}
}

\author{Gail Kaiser}
\affiliation{%
    \institution{Columbia University}
    \city{New York}
      \state{NY}
      \country{USA}
}

\author{Baishakhi Ray}
\affiliation{%
    \institution{Columbia University}
    \city{New York}
      \state{NY}
      \country{USA}
}



\begin{abstract}

Deep Learning (DL) models to analyze source code have shown immense promise during the past few years.
More recently, self-supervised pre-training has gained traction for learning generic code representations valuable for many downstream SE tasks, such as clone and bug detection.

While previous work successfully learned from different code abstractions (\eg token, AST, graph), we argue that it is also essential to factor in how developers code day-to-day for learning  general-purpose representation. 
On the one hand, human developers tend to write repetitive programs referencing existing code snippets from the current codebase or online resources (\eg Stack Overflow website) rather than implementing functions from scratch; such behaviors result in a vast number of code clones. 
In contrast, a deviant clone by mistake might trigger malicious program behaviors. 

Thus, as a proxy to incorporate developers' coding behavior into the pre-training scheme, we propose to include code clones and their deviants. 
In particular, we propose \tool, a self-supervised pre-training strategy to place benign clones closer in the representation space while moving deviants further apart. 
We show that \tool's clone-aware pre-training drastically reduces the need for expensive pre-training resources while improving the performance of downstream SE tasks. 
We also empirically demonstrate that \tool can improve existing pre-trained models to learn better representations that consequently become more efficient in both identifying semantically equivalent programs and differentiating buggy from non-buggy code.

\end{abstract}

\maketitle

\section{Introduction}
\label{sec1:intro}

Self-supervised pre-training with BERT-like models~\cite{devlin-etal-2019-bert, liu2019roberta, sanh2019distilbert, clark2020electra} (\ie a stack of Transformer-encoder layers) has achieved impressive success on many Software Engineering (SE) Tasks~\cite{guo2021graphcodebert, feng2020codebert, buratti2020cbert, cubert, xin2021syncobert, ding2021disco}. 
The main advantage of these pre-trained models is that they do not require manual labels or active supervision. 
Instead, by leveraging huge existing code corpora (\eg \textsc{Github}), these models try to capture the statistical properties of source code, and use these correlations to ``estimate'' the code properties. 
Such self-supervised pre-training aims to embed the learned estimation to code representations and consequently assists with various downstream SE tasks like clone detection, bug finding, etc. during fine-tuning.

The early work in this line directly transplants 
models and pre-training strategies from the Natural Language Processing (NLP) field to large code corpora~\cite{feng2020codebert, cubert, buratti2020cbert}. 
For example, \citet{feng2020codebert} propose CodeBERT, one of the pioneers of pre-trained code models. 
They pre-train a BERT-like model on NL-PL pairs with two token-based objectives: masked language model (MLM) and replaced token detection. 
Later researchers integrated structural properties of code into the pre-training to better understand code syntax. 
For example, GraphCodeBERT~\cite{guo2021graphcodebert} 
uses structural information such as abstract syntax trees (ASTs)~\footnote{GraphCodeBERT does not explicitly take ASTs as input, but its data flow graph is built on ASTs.} and data dependency graphs. 

\vspace{-2mm}
\paragraph{\textbf{Limitations of Existing Work.}} Since previous work on source code modeling mainly focuses on lexical or syntactic properties of code (token, AST, and graph), they successfully learn the statistical properties at the granularity of language constructs. 
However, as pointed out by \citet{hindle2012naturalness} in their seminal paper ``On the Naturalness of Software'', as well as many years ago by Donald Knuth in ``Literate programming''~\cite{knuth1984literate}, the art of coding goes beyond using programming constructs -- it is also a human experience where developers follow some day-to-day coding practices. 

For example, developers tend to introduce code clones, often by common copy-paste practices, rather than implementing functions from scratch~\cite{baker1995finding, Ducasse1999detecting, li2012cbcd, li2004cpminer}. 
Developers need to then adapt the clone to the new scope, such as reassigning the identifiers meaningful names~\cite{knuth1984literate} and adjusting control-/data-flow. 
Unfortunately, the introduction of subtle human errors is nearly unavoidable under such adaptation---\eg wrong identifiers~\cite{chou2001oserrors, li2004cpminer} and inconsistent control / data flows~\cite{ray2013detecting, juergens2009do}. 

Ignorance of common coding behaviors and likely human errors makes existing models inaccurately estimate code semantics in certain cases. 
Figure~\ref{fig:motivation_example} shows such an example collected from a real-world project: \ref{fig:original_code} and  \ref{fig:synthetic_clone} have identical functionalities, but they do not share similar tokens or structures since developers refactor it quite a bit (marked in green) after cloning. 
On the other hand, \ref{fig:original_code} and \ref{fig:clone_deviant} are syntactically very similar, yet the subtle differences in type and comparison operator (marked in red) can potentially trigger integer overflow and out-of-array access issues in the latter. 

Such examples can occur due to developers' common coding behaviors -- copy-pasting and accidentally introducing bugs, respectively.  
We refer to the former as \emph{code clones} (a.k.a. code variants) and the latter as \emph{clone deviants}. 
Ideally, the pre-trained code model should well-capture the functional similarity of programs~\cite{ding2021disco} rather than only textual overlaps, and encode \ref{fig:original_code} to be closer to \ref{fig:synthetic_clone} than \ref{fig:clone_deviant} in the representation space. 

To better understand whether existing code models encode programs as expected, we visualize the code embeddings of \{\ref{fig:original_code}, \ref{fig:synthetic_clone}, \ref{fig:clone_deviant}\} generated by GraphCodeBERT in Figure~\ref{fig:graphcodebert_pca_motivation_example}. 
Disappointingly, as we see in Figure~\ref{fig:graphcodebert_pca_motivation_example}, GraphCodeBERT ignores the functional equivalence of code, where the clone is far away from the original code. 
Also, it relies too much on syntactic similarities to represent programs, leading to the deviant and some random code in the wild being encoded closer to the original code. Other syntax-based pre-trained models operate similarly. In contrast, ~\Cref{fig:concord_pca_motivation_example} shows the desired embedding where original and its clone are closer in the representation space rather than its deviant. 


\begin{figure*}[!tp]
\begin{subfigure}{0.35\linewidth}
\begin{lstlisting}
TFStatus Eval(TfContext*
context, TfNode* node){
  TfIntArray* output_shape=...;
  const int lookup_rank=...;
  TF_LITE_ENSURE(context,...);
  int k=0;
  size_t embedding_size=1;
  size_t lookup_size=1;
  for(int i=0;i<lookup_rank-1;i++,k++)
  {
    const size_t dim=...;
    lookup_size *= dim;
    output_shape->data[k]=dim;
  }
}
\end{lstlisting}
\caption{Original Code}
\label{fig:original_code}
\end{subfigure}
\begin{subfigure}{0.28\linewidth}
\begin{lstlisting}
TFStatus @\javagreen{\textbf{model\_evaluation}}@(
TfContext* @\javagreen{\textbf{ctxt}}@, TfNode* @\javagreen{\textbf{nd}}@){
  TfIntArray* @\javagreen{\textbf{rand\_name}}@ =...;
  const int @\javagreen{\textbf{rank}}@=...;
  TF_LITE_ENSURE(@\javagreen{\textbf{ctxt}}@,...);
  @\javagreen{\textbf{size\_t lookup\_sz=1;}}@
  @\javagreen{\textbf{int z=0;}}@
  @\javagreen{\textbf{size\_t emb\_sz=1;}}@
  @\javagreen{\textbf{int x=0;}}@
  @\javagreen{\textbf{while(x<rank-1)\{}}@
    const size_t dim=...;
    @\javagreen{\textbf{lookup\_sz}}@ *= dim;
    @\javagreen{\textbf{rand\_name}}@->data[z]=dim;
    @\javagreen{\textbf{x+=1;}}@
    @\javagreen{\textbf{z+=1;}}@}}
    
\end{lstlisting}
\caption{Code Clone}
\label{fig:synthetic_clone}
\end{subfigure}
\begin{subfigure}{0.33\linewidth}
\begin{lstlisting}
TFStatus Eval(TfContext*
context, TfNode* node){
  TfeIntArray* output_shape=...;
  const int lookup_rank=...;
  TF_LITE_ENSURE(context, ...);
  int k = 0;
  @\red{\textbf{int}}@ embedding_size=1;
  @\red{\textbf{int}}@ lookup_size=1;
  for(int i=0;i@\red{\textbf{<=}}@lookup_rank-1;i++,k++)
  {
    const @\red{\textbf{int}}@ dim=...;
    lookup_size *= dim;
    output_shape->data[k]=dim;
  }
}


\end{lstlisting}
\caption{Buggy Clone-deviant}
\label{fig:clone_deviant}
\end{subfigure}

\caption{\small Motivation example: \ref{fig:original_code} is the original code; \ref{fig:synthetic_clone} is the clone of \ref{fig:original_code}; \ref{fig:clone_deviant} is the deviant clone of \ref{fig:original_code} that accidentally introduces security bugs. This example is adapted from CVE-2022-23559~\cite{cve-2022-23559} of Tensorflow~\cite{tensorflow2015-whitepaper} project.}
\label{fig:motivation_example}
\end{figure*}

\begin{figure}[!ht]
    \centering
    \begin{subfigure}[t]{0.479\columnwidth}
     \includegraphics[width=\columnwidth]{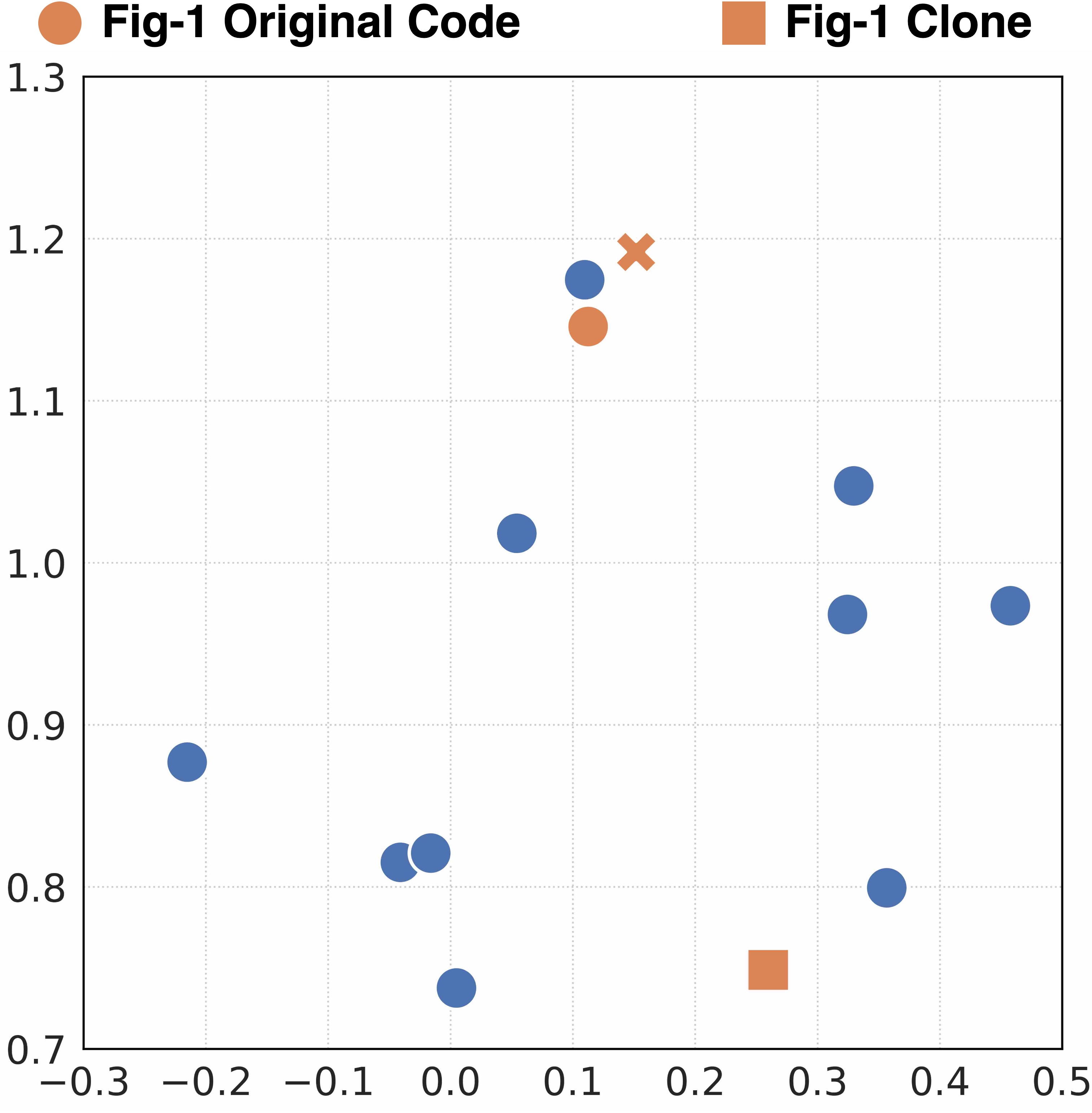}
     \caption{\small Syntax-based Code Model}
     \label{fig:graphcodebert_pca_motivation_example}
    \end{subfigure}\hfill
    \begin{subfigure}[t]{0.502\columnwidth}
     \includegraphics[width=\columnwidth]{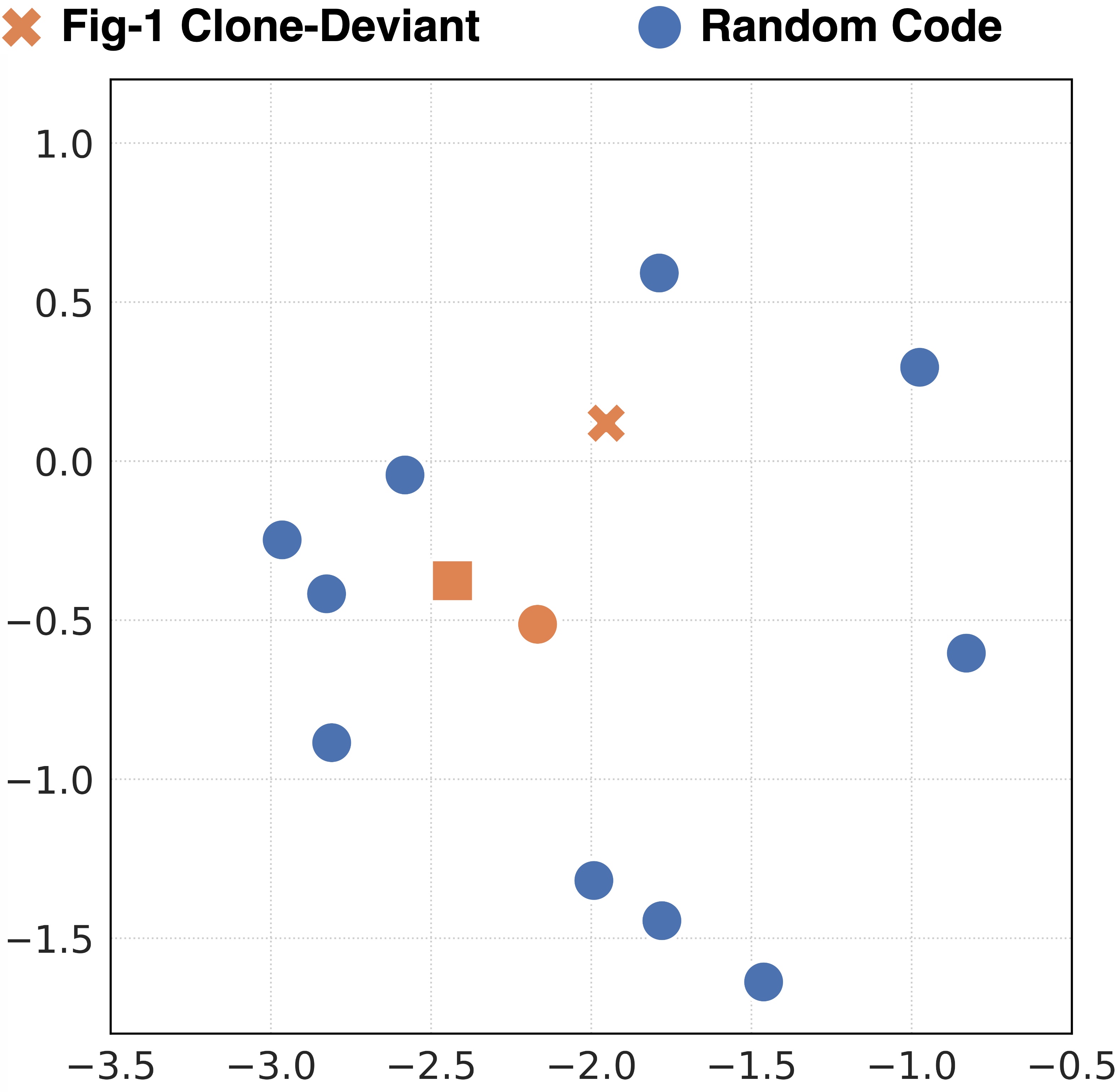}
     \caption{\small Improved by \toolbf}
     \label{fig:concord_pca_motivation_example}
    \end{subfigure}
    \caption{\small Visualization using Principle Component Analysis (PCA)~\cite{hotelling1936pca} of Figure~\ref{fig:motivation_example}'s code representations generated by GraphCodeBERT
    before (\ref{fig:graphcodebert_pca_motivation_example}) and after \toolbf's improvement (\ref{fig:concord_pca_motivation_example}).}
    \label{fig:vis_motivation}
\end{figure}

In this paper, we aim to address the limitation of existing pre-trained code models by incorporating common coding behaviors directly into the pre-training framework with a focus on  developers' code cloning practices.
However, designing such a tool is challenging because: 

\begin{enumerate}[leftmargin=*,topsep=0pt,label=\roman*., labelindent=0pt]
    \item 
    \textit{Lack of clone variants.} 
To expose diverse cloning practices commonly performed by developers, we need to train a model 
with various clone variants for each original code sample. However, manually collecting such data is almost impossible, especially at the scale of pre-training. Similarly, collecting deviants of clones is also non-trivial. 
 \item
 \textit{Learning clone functionalities.}  \tool needs to learn that although two code snippets may not be structurally similar, they still may be functionally equivalent (\eg Figures~\ref{fig:original_code} and~\ref{fig:synthetic_clone}). Similarly, clone deviants can structurally resemble each other but functionally deviate (\eg Figures~\ref{fig:original_code} and~\ref{fig:clone_deviant}). 
\end{enumerate}

\paragraph{\textbf{Our Approach.}}To address the first challenge, we propose a novel clone-aware data augmentation scheme. 
We design multiple code transformation heuristics, imitating human developers' cloning behaviors. 
Using these heuristics, we generate two variants of each program in the dataset: a) Type-1, Type-2, Type-3, and Type-4 clone, as defined by ~\cite{roy2009nearmiss, uddin2011nearmiss, Saini2021}\textemdash these are essentially a near-miss or semantic clone with equivalent functionality and b) a clone deviant with contradictory functionality (by injecting subtle bugs into the original program). 
To address the second challenge, we then learn clone functionalities with the augmented dataset.
We encode the program and its clone 
variants with similar embeddings and differentiate the buggy deviants with very distinct embeddings. 
Concretely, we pre-train a language model, \tool, with a \emph{contrastive learning} (CLR) objective~\cite{chopra2005learning, chen20simclr, gao2021simcse} to maximize the similarity between the original code and its clone 
and minimize the agreement between the original code and its clone deviant. 

Besides the CLR objective, \tool also learns the token representation of the original code using a masked language model (MLM) and the structural code properties with a new tree-structure prediction objective (LTSP) that can embed structural contexts into each token representation. 
Thus \tool is designed to learn statistical properties of PL constructs as well as developers' cloning behaviors in a single framework. 

\paragraph{\textbf{Results.}} We pre-train \tool with only 1.55 Gigabytes (GB) code\footnote{The size is measured when the code is raw text.} in a multi-lingual setting with C, C++, and Java for only 40k steps with batch size of 2048 samples during the first phase and 3k steps with batch size of 512 samples during the second phase (details in \S~\ref{subsec:pretraining-task}). 
In contrast, CodeBERT and GraphCodeBERT are pre-trained for 100k-200k steps with batch size of 2048 samples on 20GB data. 

Our results show that \tool achieves significantly better performance in clone detection and bug detection even with much cheaper training expense: \tool achieves 91.5\% MAP@R on CodeXGLUE POJ-104~\cite{msr2021codexglue, mou2016convolutional} for clone detection, outperforming the best baseline by 5.5\%, and reports the best F1 on three different bug detection benchmarks. 
We also explore \tool's extendability by adapting our approach on existing pre-trained code models. 
Our evaluation empirically shows that (1) \tool effectively improves existing
code models' performance from 82.7\% to 89.3\% in MAP@R for clone detection, from 53.1\% to 60.6\% in F1 for bug detection, and from 67.6\% to 69.7\% in MRR for code search; and (2) \tool enhances these models' code representation with semantic-aware signals, pulling the clone towards the original code representation and pushing the deviant and irrelevant programs away from it. 
Figure~\ref{fig:concord_pca_motivation_example} visualizes this enhancement.

To summarize, the main contributions of this paper are:

\begin{enumerate}[leftmargin=*,topsep=0pt]
    
    \item We design a data-augmentation technique to automatically synthesize near-miss clones~\cite{roy2009nearmiss, uddin2011nearmiss}, semantic clones and clone-deviants emulating a daily developer practice. 
    Our evaluation reveals that \tool's data augmentation is more effective and controllable than the state-of-the-art deep-learning-based augmentation strategy.
    
    \item We propose \tool, a clone-aware pre-training framework, to effectively encode program semantics into code representations. 
    Our evaluation shows that \tool outperforms existing code models in downstream SE tasks while requiring significantly less training resources (data and step size).
    
    \item We adapt our \tool approach to existing pre-trained code models and successfully improve their performance and code representation quality.
    
    \item We release \tool pre-trained model, data and code at:
    \url{https://github.com/ARiSE-Lab/CONCORD_ISSTA_23}.
\end{enumerate}

\section{Overview}
\label{sec: overview}

\begin{figure}[th]
    \centering
    \includegraphics[width=\columnwidth]{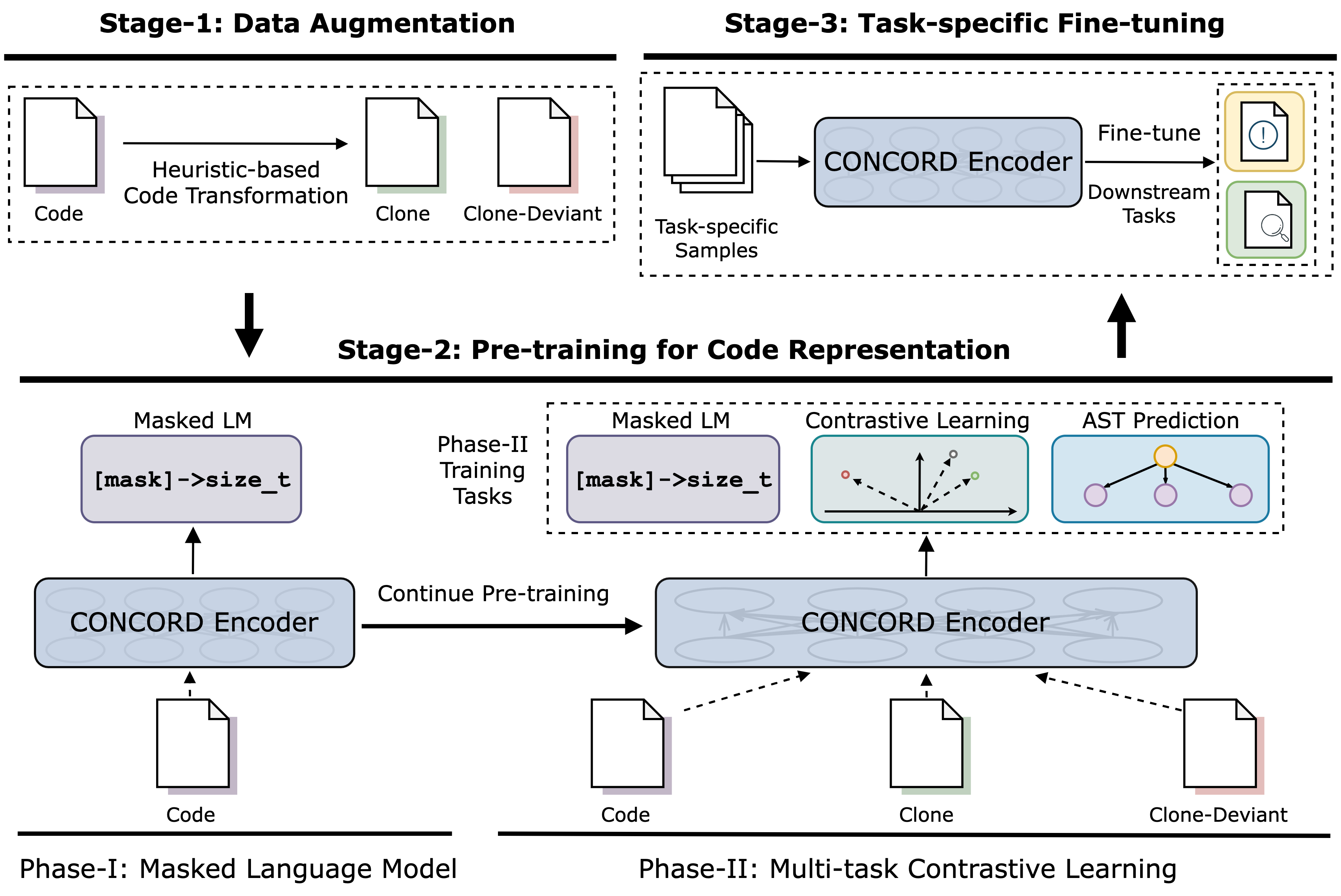}
    \caption{Overview of \toolbf.}
    \label{fig:overview}
\end{figure}

\tool is constructed based on BERT~\cite{devlin-etal-2019-bert}. Figure~\ref{fig:overview} shows the overview of \tool. \tool contains three main stages: (1) data augmentation, (2) pre-training to learn code representations, and 
(3) task-specific fine-tuning.

\noindent\textbf{Stage-1: Data Augmentation.} The goal of our data augmentation is to pair each original sample with a clone as the semantically equivalent counterpart and a clone-deviant as the buggy counterpart, so we design heuristics to augment the dataset by transforming the original code into these counterparts. The generated clone is syntactically distinct from the original sample while preserving the semantics. The generated clone-deviant shares most tokens and a similar structure to the original sample but is injected with bugs. Such augmentation forces the model to contrast code semantics rather than only syntactic properties of code during training. 

\noindent\textbf{Stage-2: Pre-training to learn code representations.} With augmented clones and clone-deviants, we pre-train \tool to learn code representations. We propose two-phase pre-training to comprehensively capture code properties, from general and statistical perspectives to structural and semantic features. 

Phase-I applies the standard masked language model training (MLM)~\cite{liu2019roberta, devlin-etal-2019-bert, feng2020codebert, guo2021graphcodebert}, where we randomly mask code tokens and train the model to predict them back. This phase teaches the model to generally understand code so that it can pick the correct token from the vocabulary given the code context. Based on such a generic model, Phase-II conducts multi-task contrastive learning to better capture the code structures and semantics. Specifically, we introduce two more objectives besides MLM: local tree-structure prediction (LTSP) and contrastive learning (CLR). LTSP trains the model to construct the abstract syntax tree (AST) given the source code, empowering it with structural knowledge of programming languages. CLR trains the model to generate code representations according to the program semantics. This objective optimizes the model toward encoding code clones with similar representations and pushing buggy deviants far from the benign code in the representation space.

Such a two-phase pre-training also leaves more flexibility for \tool, as we can adapt the multi-task contrastive learning on top of the existing code models, extending \tool to distinct tasks, and modalities. We will study such extensibility in \S~\ref{subsec:rq4_concord_variation}.

\noindent\textbf{Stage-3: Fine-tuning for Downstream Tasks.}
Finally, we load the pre-trained model as the encoder for the task-specific inputs and keep optimizing the encoder with task-specific objectives (\eg cross-entropy for classification) during the fine-tuning.

\section{\toolbf}
\label{sec:approach}

\subsection{Stage-I: Data Augmentation}
\label{subsec:data_aug}

To augment a code sample, we first represent it as AST and 
transform the AST with applicable heuristics to generate its clones and clone-deviants.

\subsubsection{Generating Clones.}
\label{subsec:synthetic_clone}

Real-world code clones are commonly classified into four categories~\cite{roy2009nearmiss, uddin2011nearmiss, Saini2021}. Type-1 defines the exact clones that programs only differ in white space and comments. Type-2 defines the clones that are syntactically similar but may contain distinct identifier names, types, and literals. Type-3 clones are still syntactically similar but consider more complicated patterns than Type-2, such as statement-level modification. 
Type-4 defines the semantic clones that programs are functionally equivalent but might not be syntactically similar. In this work, we design heuristics to generate clones, imitating the features of Type-2/3/4 clones.

\noindent\textbf{\circled{1} Identifier Renaming.}
Motivated by Type-2 clone, we design identifier (variable or function name) renaming rules to generate clones. We imitate the developers' renaming behaviors with semantic-preserving strategies: (1) If the name is a single character, such as \texttt{i}, we rename it with another single-character variable, such as \texttt{j}. (2) If the name is camel case or snake case, we will split the name into sub-words by underscores or capitals and reconstruct the name by permuting these sub-words or randomly removing part of them: for example, if we have a name called \texttt{client\_server}, we rename it as \texttt{server\_client} or \texttt{client}.

We also consider the names with more randomness. We collect a vocabulary of all possible identifier names from our datasets, and to generate clones, we will rename the identifiers in the original program with randomly selected names from this vocabulary. 

We apply the semantic-preserving first and if not successful, we apply the random renaming. We also make sure that our renaming does {\em not} change the execution behaviors of original programs.

\noindent\textbf{\circled{2} Statement Rewriting.} We design rewriting rules to synthesize Type-3 clones, imitating the developers' behaviors of writing functionally equivalent statements with varied patterns~\cite{gopstein2017understanding, gopstein2018prevalence, gopstein2020thinking}. We rewrite statements based on the three most frequent patterns: (1) We transform the conditional/ternary operators into If-Else statements. For example, we rewrite "\texttt{y = (x != 0) ? 2/x : 0;}" as "\texttt{if (x != 0) \{y = 2/x;\} else \{y = 0;\}}". (2) We rewrite the increment/decrement statements into other equivalent formats. For example, we rewrite "\texttt{y = x++;}" as "\texttt{y = x; x = x + 1;}".
(3) We mirror the comparison statements without changing the control flow. For example, "\texttt{if (x > y)}" will be rewritten as "\texttt{if (y < x)}".

\noindent\textbf{\circled{3} Block Rewriting.} To introduce more complicated Type-3 clones, we also rewrite code blocks with two main categories: (1) Loops rewriting. We design transformations to rewrite for-loop(s) as while-loop(s) and vice versa. (2) If-Else block swapping. For example, given the program "\texttt{if (a < b) \{A\} else \{B\}}", we will rewrite it as "\texttt{if (a >= b) \{B\} else \{A\}}". 

\noindent\textbf{\circled{4} Dead Code Insertion.} We create unreachable code blocks such as \texttt{if (False) \{BLOCK\}} and \texttt{while (2 < 0) \{BLOCK\}} and inject such blocks into the original code. For the statements in the \texttt{BLOCK}, we randomly pick sequential statements from the original programs and replicate them at the \texttt{BLOCK} part.

\noindent\textbf{\circled{5} Declaration/Initialization Statements Permutation.}
To implement this, we first conduct dependency analysis to identify a set of local variables that do not depend on other values for initialization. Then we move their declaration statements to the beginning of the function and permute them. For example, "\texttt{int x; int y = 0;}" will be rewritten as "\texttt{int y = 0; int x;}".

We randomly pick one or several applicable transformations to generate the clone while maintaining the program behaviors

\subsubsection{Generating Clone-deviants.}
\label{subsec:synthetic_deviants}
Clone-deviants imitate the situation that code clones accidentally introduce flaws that maliciously change the program behaviors while sharing most tokens and similar syntax with the original, benign code. We design heuristics to generate such deviants based on the observations from a wide number of reported, real-world bug patterns~\cite{mitre2020cwe, mitre2022cve, ding2020patching, allamanis2021self, chakraborty2021reveal, karampatsis2020bigcode, just2014defects4j, ding2021disco}.

\noindent\textbf{\circled{1} Operator Bugs.} We randomly replace the comparison operator with another of the same type to change the control flow and replace arithmetic operator to trigger unexpected program behaviors.

\noindent\textbf{\circled{2} Data-type Bugs.}
Wrong data types can trigger several security flaws (\eg integer overflow).  
We intentionally change the data types in the original code to inject such bugs, while ensuring the new code can still be compiled.

\noindent\textbf{\circled{3} Variable Bugs.} 
We induce such bugs with two approaches: (1) we perform scope analysis on the AST and replace a variable with another unexpected variable reachable in the same scope. (2) we remove the initialization expression of variables.

\noindent\textbf{\circled{4} Value Bugs.}
We inject bugs by replacing a Boolean value with its opponent and an arithmetic value with random numbers.

\noindent\textbf{\circled{5} Pointer Bugs.}
To inject such bugs, we randomly remove the initialization expression during pointer declaration or set some pointers to {\tt NULL}.

\noindent\textbf{\circled{6} Statement Bugs.} We randomly remove small condition-checking statements/blocks, which are
typically used to check necessary preconditions before doing critical operations (\eg checking the index's validity before accessing an array)

\noindent\textbf{\circled{7} Function-call Bugs.} 
For a randomly selected function call, we add, remove, swap or assign {\tt NULL} value to arguments, forcing the code to behave unexpectedly. 

\subsection{Stage-II: Pre-training}
\label{subsec:pretraining-task}

\subsubsection{Input Representation.}
\label{subsubsec:input_repr}
As shown in Figure~\ref{fig:mlm}, given the program,
we parse it and flatten it as a sequence of code tokens $S = \{s_1, ..., s_m\}$.
To alleviate the out-of-vocabulary concern of programming languages~\cite{karampatsis2020bigcode}, we train a SentencePiece~\citep{kudo-richardson-2018-sentencepiece} subword tokenizer based on such flattened code token sequences with vocabulary size of 50,000. 
We use this tokenizer to divide $m$ source code tokens into $k$ sub-tokens ($m \leq k$). Similar to BERT, we prepend the special token {\tt [CLS]} and append the special token {\tt [SEP]} to  the sub-token sequence $C = \{[CLS], c_1, ..., c_k, [SEP]\}$.
Finally, \tool converts the pre-processed code sequence to vectors $V = \{v_{[CLS]}, v_1, ..., v_k, v_{[SEP]}\}$ with a token embedding layer. 


\subsubsection{Phase-I: Learning Code Texts.}
\label{subsubsec:learn_code_text} 
\begin{figure}[t]
    \centering
    \begin{subfigure}[t]{0.58\columnwidth}
     \includegraphics[width=\columnwidth]{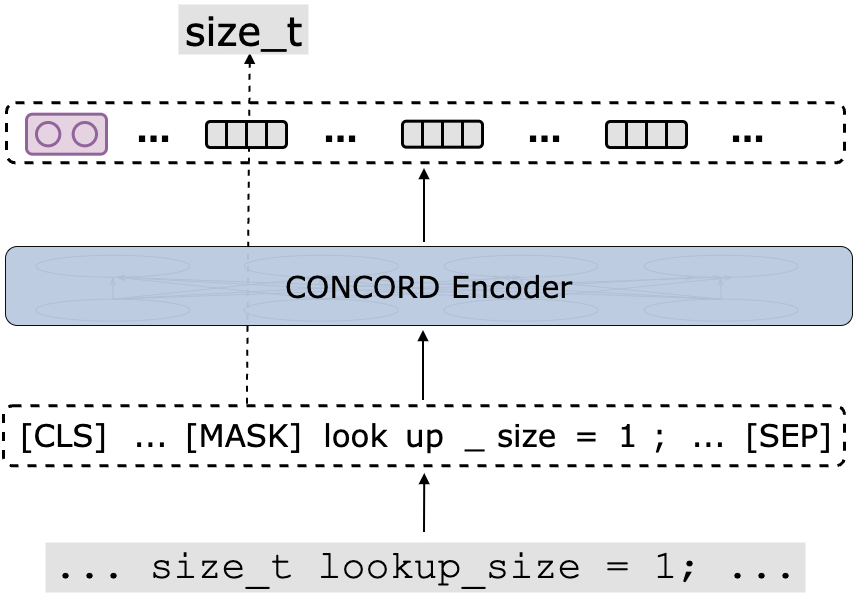}
     \caption{\small Input Representation \& MLM}
     \label{fig:mlm}
    \end{subfigure}\hfill  
    \begin{subfigure}[t]{0.4\columnwidth}
     \includegraphics[width=\columnwidth]{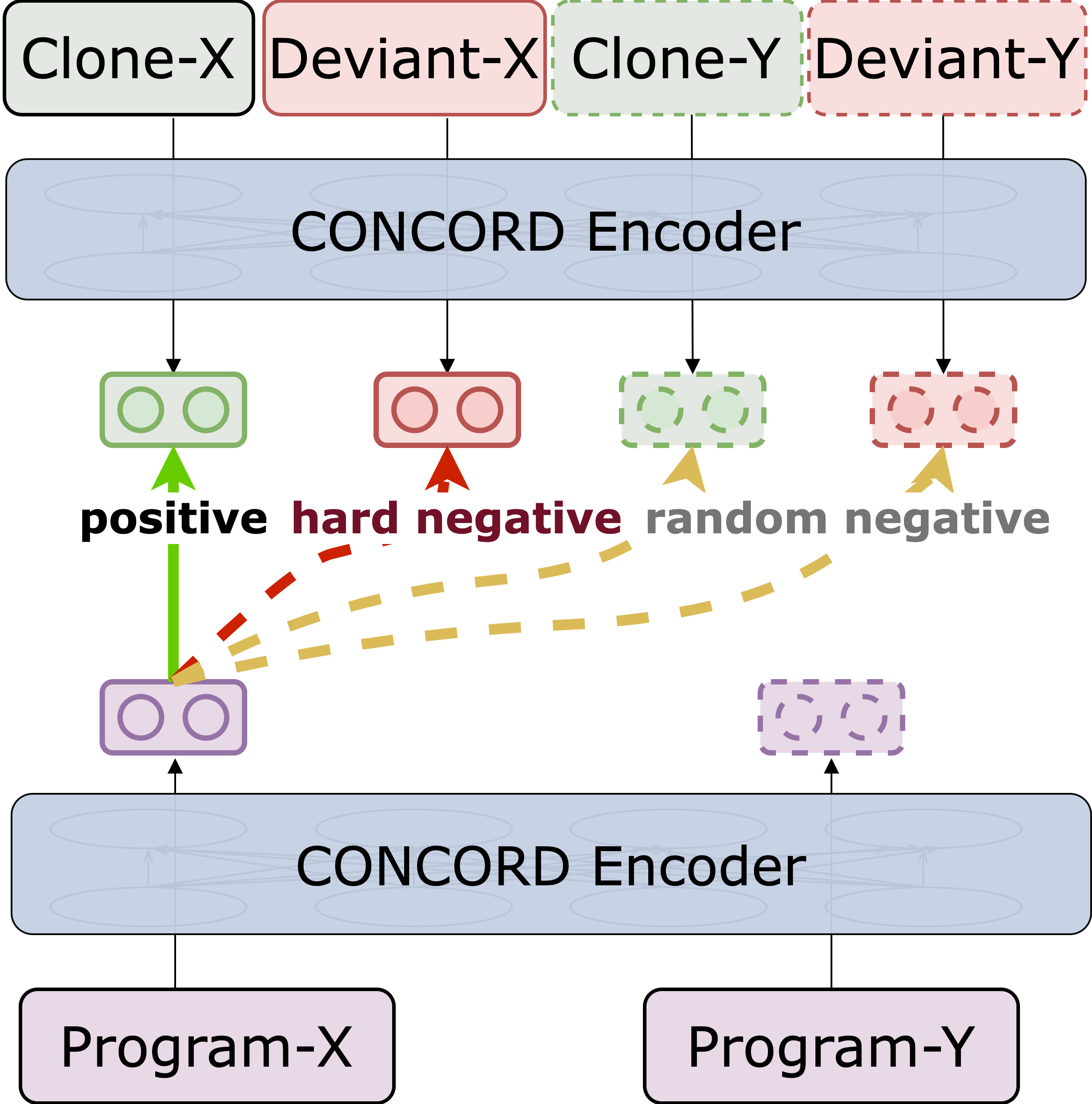}
     \caption{\small Contrastive Learning}
     \label{fig:clr}
    \end{subfigure}
    \caption{\small Details of Pre-training Tasks}
    \label{fig:mlm_and_clr}
\end{figure}

In Phase-I, we pre-train the model with masked language model (MLM)~\cite{devlin-etal-2019-bert} to capture the \emph{naturalness}~\cite{hindle2012naturalness, ray2016naturalness} of code text (Figure~\ref{fig:mlm}). Concretely, given the code sequence $C$, we randomly choose 15\% of tokens (\eg \texttt{size\_t} in~\ref{fig:mlm}) and replace them with a special token \texttt{[MASK]}. We denote the set of masked token index as $M$, and the masked tokens as $\{c_m | m \in M\}$. The model will learn to encode the surrounding contexts (\eg \texttt{lookup\_size=1} in~\ref{fig:mlm}) of \texttt{[MASK]}s into their hidden states output by the model $\{h_m | m \in M\}$, and reconstruct the masked tokens conditioned on them. We compute the negative log likelihood of the masked tokens as the loss for this phase.
\begin{equation}
\label{eq:mlm_loss}
    \mathcal{L}_{MLM} = \sum_{m \in M} - log P(c_m | h_m)
\end{equation}
Equation~\ref{eq:mlm_loss} optimizes the model to minimize this negative loss during training, which maximizes the likelihood of original tokens before masking, given their surrounding contexts. It guides the model weights to fit the source code distribution in the wild. Consequently, the trained model will be able to produce code representations following such general distribution.

\subsubsection{Phase-II: Learning Code Structures and Semantics.}
\label{subsubsec:learn_code_structure_and_functionality}
In Phase-II, we load the model weights from Phase-I and continue to learn the syntactic and semantic perspectives of programs.

\begin{figure}[t]
    \centering
    \includegraphics[width=0.8\columnwidth]{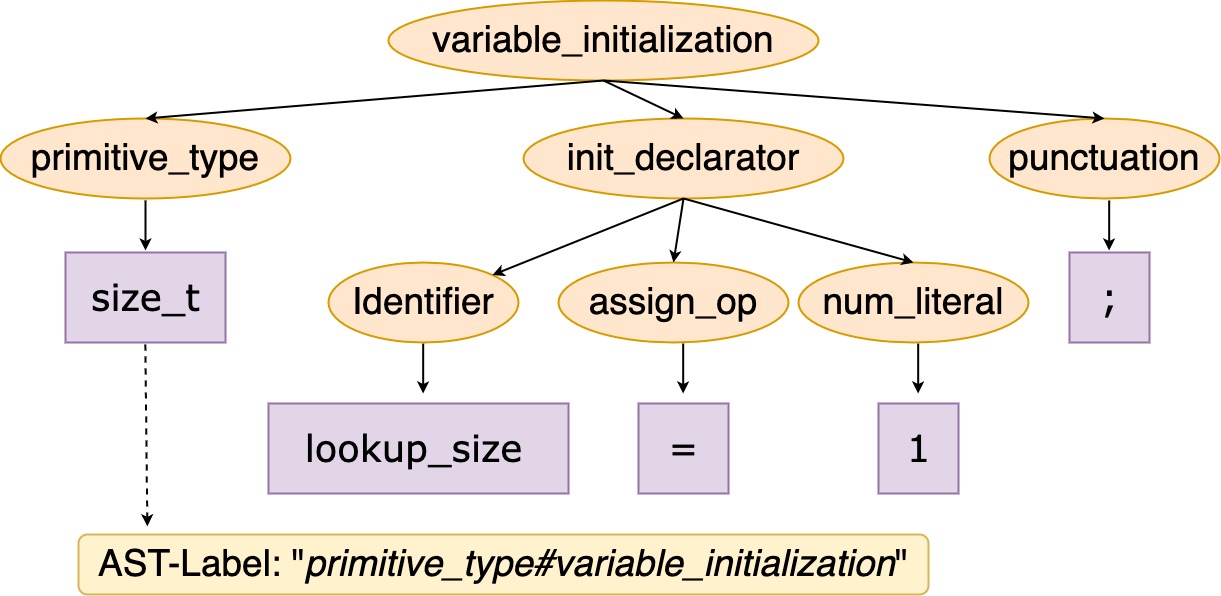}
    \caption{\small Building AST-Label for LTSP.}
    \label{fig:AST_label}
\end{figure}

\noindent\textbf{Local Tree Structure Prediction (LTSP)}.
To learn the code structure, we propose LTSP, teaching the model to predict the local ASTs given the code text. Concretely, we assign every code token a local AST label (Figure~\ref{fig:AST_label}), $tt\#pt$, comprising of the type of the corresponding terminal node ($tt$) (\eg keyword, identifier), and the type of the immediate parent node ($pt$)(\eg for-statement, declaration). For example, in Figure~\ref{fig:AST_label}, the token \texttt{size\_t} is a primitive type in the variable initialization statement, so it will have a AST label of \textit{primitive\_type\#variable\_initialization}. 
All sub-tokens of a token will share the same label. Essentially, we are encoding the information of a 2-layer sub-tree into the AST-Label, and with such labels, the model can comprehensively capture the local dependencies, such as the connection with parent, children, and sibling nodes. We parse our dataset and exhaustively build the AST-Label vocabulary with all possible labels.

Formally, we pre-define the AST-Label sequence by parsing the code sequence, $T = \{[CLS], t_1, t_2, ...,$ $t_k, [SEP]\}$, and we use this sequence as the ground-truth of the LTSP task. Similar to Phase-I, the model input is just the source code sequence ($C$), and the Transformer encoder will output a representation $h_i$ for $c_i \in C$, and we train the model to predict the local AST type for \emph{every} token based on $h_i$. We present the loss for LTSP as 

\begin{equation}
  \mathcal{L}_{LTSP} = \sum_{i} - log P(t_i | h_i)  
\end{equation}

Compared with other structure-based models~\cite{guo2021graphcodebert, Jiang2021TreeBERT, ding2021disco}, \tool has the advantage of not requiring structure input during the pre-traing and fine-tuning, but is aware of code structures and language grammars.

\noindent\textbf{Contrastive Learning w/ Clones and Deviants.} \revision{Contrastive learning has been proven to be effective in learning the semantic similarity of source code~\cite{ding2021disco, bui2021corder, guo2022unixcoder, xin2021syncobert, jain2020contrastive}. It is realized by optimization functions that maximize the representation similarity among semantically equivalent programs and enlarge the distances among semantically distinct or irrelevant code. We apply the contrastive learning objective to encode the clone-aware semantic signals into the code representations.}

As shown in Figure~\ref{fig:clr}, for each program in the dataset (e.g., Program-X), the data augmentation generator (\S\ref{subsec:data_aug}) creates a clone (Clone-X) as the positive counterpart and a clone-deviant (Deviant-X) as the negative counterpart. For each batch (Program-X, Y is one batch), \tool builds one positive pair (green arrow), one hard-negative pair (red, dashed arrow) and several random negative pairs using other in-batch samples (yellow, dashed arrows). Given all these pairs, we train the model to maximize the cosine similarity of the positive pair's representations and minimize the similarity of negative pairs. Formally, we have a minibatch of $N$ programs, and for each program, we use the encoder output of \texttt{[CLS]} token to represent the whole sequence: $\mathbf{z} = h_{[CLS]}$. With the data augmentation, the minibatch is extended to N triplets of $(\mathbf{z}, \mathbf{z}^+, \mathbf{z}^-)$, where $\mathbf{z}^+$ corresponds to the generated clone, and $\mathbf{z}^-$ corresponds to the clone-deviant.
We refer to the contrastive loss with hard negative samples from \citet{gao2021simcse} and we adapt it to our scope as follows.
\begin{equation}
    \label{eq:clr_objective}
    \mathcal{L}_{CLR} = -\log\frac{\mathbf{e}^{\text{sim}\left(\mathbf{z}, \mathbf{z}^+\right)/\tau}}{\sum^{N}_{n=1}\left(\mathbf{e}^{\text{sim}\left(\mathbf{z}, \mathbf{z}_{n}^+\right)/\tau} + \mathbf{e}^{\text{sim}\left(\mathbf{z}, \mathbf{z}_{n}^-\right)/\tau} \right)}
\end{equation}

In equation~\ref{eq:clr_objective}, we use cosine similarity as the $sim()$ function and $\tau$ is a parameter to scale the loss, and similar to~\cite{gao2021simcse} we use $\tau = 0.05$. 

Similar to existing works~\cite{feng2020codebert, guo2021graphcodebert}, we keep learning the code text using MLM during the second phase of pre-training, together with LTSP and contrastive loss. Therefore, the final loss function to optimize Phase-II pre-training is as follows, where $\lambda_1=1.0$, $\lambda_2=0.1$, $\lambda_3=1.0$ respectively.
\begin{equation}
    \mathcal{L}(\theta) = \lambda_1\cdot\mathcal{L}_{MLM}(\theta) + \lambda_2\cdot\mathcal{L}_{LTSP}(\theta) + \lambda_3\cdot\mathcal{L}_{CLR}(\theta)
\end{equation}

\subsection{Stage-III: Fine-tuning}
We apply the standard transfer learning strategy to the pre-trained model for concrete downstream tasks: we load the pre-trained model as the encoder to generate generic code representations and keep optimizing the model with supervised fine-tuning. We consider semantic-clone detection and bug detection as \tool's main applications. 

\noindent\textbf{Semantic Clone Detection.} Detecting semantic clones is significant for software maintenance~\cite{kim2017vuddy,li2016vulpecker} yet very challenging in practice, since the token overlap among semantic clones is quite limited and syntactic structures are not similar as well. This task evaluates the model's capacity of retrieving semantic clones: given a program as query and a set of random programs as candidates, the model needs to identify the semantic clones of the query out of thousands of candidates.

\noindent\textbf{Bug Detection.} Human errors are the main causes for software flaws. For example, when developers adapt clones into a new scope, small errors, such as wrong identifier names and operators, are accidentally introduced~\cite{li2004cpminer, ray2013detecting}, and such bugs are only a few tokens away from the benign version. Similarly, it is challenging for models to identify bug-fixing patches as benign, if developers only repair a few tokens ~\cite{ding2020patching}. Without attending to such human behaviors, models struggle with false positives and false negatives~\cite{pradel2018deepbug, Kharkar2022LearningTR}. \tool alleviates such concerns by incorporating these behaviors into the pre-training, synthesizing clone-deviants as the hard negative samples and forcing the model to differentiate bugs and benign code that are syntactically similar.

We evaluate the model's capacity of detecting software bugs at function-level. Specifically, given a code function as input, the model needs to classify it as buggy (positive) or benign (negative).

\section{Experimental Setup}
\label{subsec:expr_setup}

\subsection{Pre-training Dataset}
 
We collect our pre-training corpus from open-source GitHub projects. We rank Github repositories by the number of stars and focus on the most popular ones. 
After filtering out forks from existing repositories,
we collect the dataset for each language from top-100 repositories. We only consider the ``.c'', ``.cpp'' and ``.java'' files for C, C++, and Java repositories respectively. Similar to the existing datasets~\cite{husain2019codesearchnet}, we extract the code functions/methods from the code files. The raw datasets for C, C++, and Java are of size 662MB, 330MB, and 556MB respectively.  
We use the state-of-the-art, multi-lingual AST parser, Tree-sitter~\cite{treesitter} to parse the source code.

\subsection{Model Configuration}

\tool applies a standard BERT\textsubscript{BASE} architecture~\cite{devlin-etal-2019-bert} with 12 layers of Transformer-encoder, and each layer has 12 attention heads and the hidden dimension is 768. \revision{The maximum sequence length is 512 BPE tokens, and the longer sequence will be truncated. As samples in our datasets are function-level code, truncation does not frequently happen.} Our experiments are conducted on 2 $\times$ 24GB NVIDIA GeForce RTX-3090 GPUs. For Phase-I, we pre-train \tool for 40k batches with a batch size of 2048, and we use the learning rate of 5e-4. \revision{1e-4 to 5e-4 is the common range for learning rates of MLM-based code models~\cite{feng2020codebert, cubert, buratti2020cbert, ding2021disco, guo2021graphcodebert}, and we follow~\cite{feng2020codebert} to select 5e-4}. For phase-II, we further pre-train \tool for 3k batches with a batch size of 512, and we use the learning rate of 5e-5. For all the fine-tuning tasks, \tool uses the learning rate of 8e-6. \revision{Learning rates typically decrease for later phases~\cite{feng2020codebert, guo2021graphcodebert, ding2021disco}, so \tool follows the same design}. We use Adam optimizer~\cite{Kingma2015AdamAM} with the linear learning rate decay. \tool is implemented mainly with Pytorch~\cite{paszke2019pytorch} and Huggingface~\cite{wolf-etal-2020-huggingface} libraries.

\subsection{Evaluation Datasets \& Metrics}

\noindent\textbf{Semantic Clone Datasets.} We consider two datasets for the semantic clone detection: CodeXGLUE-POJ104~\cite{mou2016convolutional, msr2021codexglue} and CodeNet-Java250~\cite{puri2021codenet}. CodeXGLUE-POJ104 contains 104 programming challenges, and each has 500 C/C++ solutions submitted by different students. CodeXGLUE~\cite{msr2021codexglue} reconstruct it as a public benchmark by splitting the dataset into Train (64 challenges), Dev (16 challenges), and Test (24 challenges) sets, making sure that there is no overlapped challenges between any two sets. 
CodeNet-Java250 contains 250 Java programming challenges from online judge websites, and each has 300 solutions from programmers. It splits the datasets into Train (125 challenges), Valid (62 challenges), and Test (63 challenges) without overlapped challenges. The detailed statistics of these two datasets can be found in Table~\ref{tab: downstream_dataset}. 

We notice that some existing works are evaluated on CodeXGLUE-BigCloneBench~\cite{svajlenko2014towards}. We do not use this benchmark, because we find that the labels are rather noisy and inaccurate~\footnote{Corresponding GitHub issue: \url{https://github.com/microsoft/CodeXGLUE/issues/93}}\footnote{Corresponding GitHub issue: \url{https://github.com/microsoft/CodeXGLUE/issues/99}}.

\noindent\textbf{Metrics of Clone Detection.} Both CodeXGLUE~\cite{msr2021codexglue} and CodeNet~\cite{puri2021codenet} are using MAP@R (Mean Average Precision @ R)\footnote{\url{https://en.wikipedia.org/wiki/Evaluation_measures_(information_retrieval)\#Mean_average_precision}}, so we follow such a design. Average precision at R is a common metric to evaluate the quality of information retrieval; it measures the average precision scores of a set of the top-R clone candidates presented in response to a query program. The "R" for CodeXGLUE is 499 as it has 500 solutions for each challenge, and we do not consider the code itself as a clone, and for CodeNet is 299, since it has 300 solutions for each problem.

\begin{table}[h]
\centering
\caption{\small Details of downstream tasks datasets.}\label{tab: downstream_dataset}
\small
\vspace{-2mm}
\begin{tabular}{l|c|c|r|r|r}\hline
Task &Dataset &Lang. &Train &Valid &Test \\\hline
\multirow{2}{*}{Clone Detection} &CXG-POJ104 &C/C++ &32,000 &8,000 &12,000 \\
&CN-Java250 &Java &37,500 &18,600 &18,900 \\\hline
\multirow{3}{*}{Bug Detection} &REVEAL &C/C++ &15,867 &2,268 &4,535 \\
&D2A &C/C++ &4,644 &597 &619 \\
&CXG-Devign &C/C++ &21,854 &2,732 &2,732 \\\hline
\end{tabular}
\end{table}

\noindent\textbf{Bug Datasets.}
We choose three public datasets: REVEAL (RV)~\cite{chakraborty2021reveal}, D2A~\cite{zheng2021d2a}, and CodeXGLUE-Devign (CXG-DV)~\cite{msr2021codexglue, Zhou2019DevignEV}.
Chakraborty \etal curated REVEAL, imitating the real-world scenario that bugs are always rare compared to the normal programs, so it ends up with the ratio of the buggy to benign samples being roughly 1:10.
D2A is a balanced dataset focusing on bug-fixing commits and annotates the previous version of modified functions as buggy and the fixed version as benign. 
CodeXGLUE-Devign is another balanced dataset introduced by \citet{Zhou2019DevignEV}, and CodeXGLUE reconstructs the dataset as a public benchmark, fixing the train/valid/test splits so that all models can be evaluated with the same splits. 

\noindent\textbf{Metrics of Bug Detection.} 
REVEAL is a super imbalanced dataset, so following the design of the original paper, we use Precision, Recall, and F1 as the evaluation metrics. D2A and Devign are balanced datasets, so we use Accuracy and F1.
\section{Evaluation}
\label{sec:evaluation}

\tool aims to learn more meaningful code representations by incorporating common coding behaviors directly into the pre-training framework, focusing on code clones and bugs introduced by these behaviors. In this section, we present the evaluation results and analysis. In particular, we ask the following \revision{five} RQs:

\begin{itemize}[leftmargin=*]
    \item {\textbf{RQ1:}} How effective is \tool \wrt state-of-the-art baselines for (1) clone detection and (2) bug finding?
    \item {\textbf{RQ2:}} How effective is \tool's data augmentation \wrt state-of-the-art deep-learning-based data augmentation?
    \item {\textbf{RQ3:}} \revision{Can \tool's LTSP pre-training objective help to learn better code presentations with code structures?}
    \item {\textbf{RQ4:}} Can \tool's pre-training improve existing code models for downstream tasks?
    \item {\textbf{RQ5:}} Can \tool learn more meaningful representations and better identify semantic similarity of programs than existing code models?
\end{itemize}

\subsection{RQ1.~Comparing \toolbf to Baselines}
\label{subsec:rq1}
\subsubsection{RQ1-A.~Semantic Clone Detection.} We present the baselines and results of semantic clone detection in this section.

\noindent\textbf{Baselines.}
We choose the best-performing pre-trained models reported by CodeXGLUE-POJ104~\cite{msr2021codexglue}
: RoBERTa~\cite{liu2019roberta}, CodeBERT~\cite{feng2020codebert}, and GraphCodeBERT~\cite{guo2021graphcodebert}. These pre-trained models have already been proven to be more effective than previous work~\cite{msr2021codexglue}, such as SourcererCC~\cite{sajnani2016sourcerercc} and Aroma~\cite{luan2019aroma}, so we no more include these older approaches in our results. We also consider two contrastive-learning-based code models: Corder~\cite{bui2021corder}\footnote{Corder has several variants, and we reported Corder-Transformer results according to the original paper, as \tool applies the same model architecture.} and  DISCO~\cite{ding2021disco}\footnote{For a fair comparison, we re-implement and pre-train DISCO with the same setting as CONCORD (\eg library versions). The evaluation shows the new implementation slightly improves the original version~\cite{ding2021disco} with acceptable error bound.}.  
We directly compare the originally reported results from the paper (\eg Corder) or the benchmark (\eg CodeBERT and RoBERTa), if available. Otherwise, we will fine-tune the baselines.

\begin{table}[!htp]\centering
\caption{\small MAP@R (\%) Results of Semantic-clone Detection}\label{tab:clone_detection}
\small
\vspace{-1mm}
\resizebox{\columnwidth}{!}{
\begin{tabular}{l|c|c|c}
\hlineB{2}
\textbf{Models} & \textbf{Data Size}&\textbf{CodeXGlue-POJ104} &\textbf{CodeNet-J250} \\
\hlineB{2}
\rowcolor{rowgray}
Corder$^*$ & 1.9 GB & 72.0 & -\\

DISCO & 1.8 GB & \revision{82.5} & \revision{76.5}\\\hline
\rowcolor{rowgray}
Corder$^*$ & 9.3 GB &84.1 & -\\

RoBERTa & 160 GB &76.7 &75.5 \\
\rowcolor{rowgray}
CodeBERT & 20 GB &82.7 &81.1 \\

GraphCodeBERT & 20 GB &86.7 &84.3 \\


\hlineB{2}
\rowcolor{rowgray}
\tool & \textbf{1.5 GB} &\textbf{91.5} &\textbf{86.5} \\
\hlineB{2}
\end{tabular}}
\begin{flushleft}
\small
\revision{*Corder applies a slightly different setting from the commonly used CodeXGLUE benchmark on POJ-104: it randomly samples 50 programs for each problem in POJ-104, resulting in 5,200 samples, and reports the mean average precision (MAP). In the strict Corder's setup, \tool reports 90.4 MAP in POJ-104.}
\end{flushleft}
\end{table}

\noindent
\textbf{Results.} The results are shown in Table~\ref{tab:clone_detection}. \tool achieves \textbf{91.5\%} MAP@R for CodeXGLUE-POJ104 and \textbf{86.5\%} for CodeNet-Java250. 
Even if pre-trained with a very small dataset (7.5\% size of GraphCodeBERT's dataset), \tool is still a clear winner with a significant margin. Comparing \tool with syntax-based baselines (\ie CodeBERT, GraphCodeBERT, and RoBERTa), the results reveal the effectiveness of contrastive learning: learning to contrast code functionalities can improve the model's performance while reducing the training cost. Interestingly, we found that \tool can also outperform other contrastive-learning-based approaches, which empirically proves the effectiveness of \tool's data augmentation and multi-task pre-training. We will conceptually discuss the different design options among these contrastive-learning-based code models in 
Related Work (\S~\ref{sec2:related})

\subsubsection{RQ1-B.Bug Detection.}
\label{subsec: bug_detect}
We present the baselines and results of bug detection in this section.

\noindent\textbf{Baselines.}
We compare \tool with Transformer-based pre-trained models containing a similar number of Transformer layers (12 layers in total), since Transformer models with more layers always significantly outperform those with fewer parameters~\cite{devlin-etal-2019-bert, liu2019roberta, gpt3, wang2021codet5}. Thus, we consider either 12-Layer Transformer-Encoder models or 6-Layer Transformer-Encoder-Decoder models.: 
Again, for released pre-trained models, we conduct full experiments on all three benchmarks, and for the others, we directly take their reported results in the original paper. 

\begin{table}[!htp]\centering
\caption{\small Results of Bug Detection.}\label{tab: bug_detection}
\small
\vspace{-2mm}
\resizebox{\linewidth}{!}
{\begin{tabular}{l|c|r|r|r|r|r|r|r}
\hlineB{2}
\multirow{2}{*}{\textbf{Model}}&\textbf{Data}&\multicolumn{3}{c|}{\textbf{RV}} &\multicolumn{2}{c|}{\textbf{D2A}} &\multicolumn{2}{c}{\textbf{CXG-DV}} \\\cline{3-9}
  &\textbf{Size}&\textbf{P.} &\textbf{R.} &\textbf{F1.} &\textbf{Acc.} &\textbf{F1.} &\textbf{Acc.} &\textbf{F1.} \\
\hlineB{2}
\rowcolor{rowgray}
\revision{DISCO} & \revision{1.8 GB} &\revision{47.9} &\revision{46.4} &\revision{47.2} &\revision{60.2} &\revision{57.9} &\textbf{64.2}&\textbf{58.5}\\

CodeBERT & 20 GB  &47.0 &47.7 &47.3 &59.2 &63.6 &63.4 &53.1 \\
\rowcolor{rowgray}
GraphCodeBERT & 20 GB  &\textbf{55.9} &39.9 &46.6 &61.0 &66.1 &62.9 &56.3 \\

PLBART & 576 GB  &44.9 &41.6 &43.2 &57.0 &61.2 &62.5 &57.9 \\
\rowcolor{rowgray}
CodeT5 & >20 GB &47.1  &46.7 &46.9 &58.9 &56.1 &62.8 &58.3 \\ 
\hlineB{2}
\rowcolor{rowgray}
\tool & \textbf{1.5 GB} & 47.8& \textbf{49.3}& \textbf{48.6}& \textbf{62.1}& \textbf{67.1}& 63.7& 58.3 \\
\hlineB{2}
\end{tabular}}
\end{table}

\noindent
\textbf{Results.} From Table~\ref{tab: bug_detection}, we can see that, pre-trained with clone-aware signals, \tool is effective at reducing the false positives and false negatives: \tool reported better F1 for all three benchmarks than those pre-trained models focusing on code syntax. In particular, when all the baselines are reporting many false negatives in REVEAL, due to the rareness of positive samples in it, \tool achieves significantly higher recall than competing models, even if these baselines are pre-trained with significantly more samples. This result empirically reveals that \tool's augmented deviants help the model reduce the confusion when differentiating buggy from benign code, even if they are sometimes syntactically similar. Interestingly, we notice DISCO~\cite{ding2021disco} performs slightly better in the CodeXGLUE benchmark, which also demonstrates the effectiveness of contrastive learning for code. However, \tool outperforms DISCO in all other bug detection benchmarks as well as the clone detection tasks, proving \tool is in general more effective at learning code semantics than DISCO.

\RS{1}{Even if pre-trained with \textbf{significantly less} data, \tool \textbf{outperforms} the state-of-the-art baselines that are not trained with the clone awareness in downstream tasks.}

\vspace{-2mm}
\subsection{RQ2. Effectiveness of \toolbf's data augmentation}

Data augmentation is the key to the success of contrastive learning models~\cite{chen20simclr, gao2021simcse}. It integrates strong bias regarding the semantic similarity of data into the model, since, during training, the model learns to maximize/minimize the similarity of the original sample with its positive/negative counterparts, which are completely generated by the pre-defined data augmentation strategy. In this section, we compare \tool's data augmentation with the state-of-the-art deep-learning-based approach.

The state-of-the-art deep-learning-based data augmentation~\cite{gao2021simcse, guo2022unixcoder} for code relies on the model's randomness to generate positive/negative counterparts implicitly, which is difficult to control and interfere with domain knowledge. Differently, \tool proposes to explicitly augment code datasets with carefully crafted heuristics (\S~\ref{subsec:data_aug}), by imitating the developers' behaviors of cloning existing code.

\noindent\textbf{Baseline and Setup}.  SimCSE~\cite{gao2021simcse} is the state-of-the-art contrastive learning framework for natural languages, and it has been proven to be effective in learning better code representations~\cite{guo2022unixcoder}. SimCSE leverages ``dropout''~\cite{srivastava2014dropout}, a deep-learning technique initially proposed to avoid model over-fitting, to generate positive samples. Concretely, the dropout mechanism randomly disables certain neurons in the neural network following Bernoulli distribution, and the randomness of each neuron is independently seeded. Therefore, SimCSE passes the \emph{same} sample twice through the model and with dropout, it will get two slightly different embeddings, which will be regarded as semantic equivalent pair. SimCSE builds negative samples using randomly sampled data points within the same batch. 

To conduct the comparison, we augment the original dataset with SimCSE data augmentation rather than \tool's, and re-train the model with the SimCSE-augmented dataset. We report results from the original \tool and \tool-SimCSE (\ie SimCSE for short in Table~\ref{tab: rq_data_aug_comp}) on downstream tasks. 

\begin{table}[!htp]\centering
\caption{\small Performance of \toolbf with SimCSE-augmented dataset and \toolbf-augmented dataset.}\label{tab: rq_data_aug_comp}
\small
\resizebox{\linewidth}{!}
{
\begin{tabular}{l|c|c|c|c|c|c|c|c|c}
\hlineB{2}
\textbf{Task} &\multicolumn{2}{c|}{\textbf{Clone Det.}} &\multicolumn{7}{c}{\textbf{Bug Detection}} \\\hline
\textbf{Dataset} &\textbf{P104} &\textbf{J250} &\multicolumn{3}{c|}{\textbf{RV}} &\multicolumn{2}{c|}{\textbf{D2A}} &\multicolumn{2}{c}{\textbf{CXG-DV}} \\
\hlineB{2}
\textbf{Metric} &\multicolumn{2}{c|}{\textbf{MAP@R}} &\textbf{P.} &\textbf{R.} &\textbf{F1} &\textbf{Acc} &\textbf{F1} &\textbf{Acc} &\textbf{F1} \\
\hlineB{2}
SimCSE & 88.5& \textbf{86.5}& \textbf{50.9}& 46.5& \textbf{48.6}& 61.3& 67.0& 63.6& 56.3\\\hlineB{2}
\rowcolor{rowgray}
CONCORD & \textbf{91.5}& \textbf{86.5}& 47.8& \textbf{49.3}& \textbf{48.6}& \textbf{62.1}& \textbf{67.1}& \textbf{63.7}& \textbf{58.3}\\
\hlineB{2}
\end{tabular}}
\end{table}

\noindent\textbf{Results and Analysis.}~The results are shown in Table~\ref{tab: rq_data_aug_comp}. \revision{\tool performs marginally better than \tool-SimCSE. Besides, CONCORD’s augmentation is easily controlled by heuristic designs while SimCSE completely relies on dropout randomness. CONCORD’s clone-aware augmentation is a proof-of-concept of integrating developers’ cloning behaviors into code representation and using the same philosophy, we could propose heuristics that align with other human requests, while SimCSE does not have such flexibility.}

\RS{2}{\revision{\tool's data augmentation proposes carefully crafted heuristics to imitate developers’
cloning patterns and bugs, and it reports comparable performance with state-of-the-art deep-learning-based data augmentation in clone detection and bug finding tasks.}}

\subsection{\revision{RQ3. Effectiveness of \toolbf's LTSP pre-training objective}}

\revision{\tool proposes a new pre-training objective, LTSP, to guide the model to learn the code syntax during the pre-training. In this RQ, we study the effectiveness of this new pre-training objective.

\noindent\textbf{Setup.} To conduct a strict comparison, we pre-train a \tool variant by removing the LSTP objective but keeping all other settings the same as the main model. As we mentioned in \S~\ref{subsubsec:learn_code_structure_and_functionality}, LTSP objective 
does not require any parsing or pre-processing on the source code during the fine-tuning for downstream tasks. Therefore, we evaluate \tool-without-LTSP using exactly the same fine-tuning data and strategies as discussed in \S~\ref{subsec:rq1} and compare its performance with the main model.
}

\begin{table}[!htp]\centering
\caption{\small The comparison of \toolbf's performance between with and without LTSP objective during pre-training.}\label{tab: ltsp_ablation}
\small
\resizebox{\linewidth}{!}
{
\begin{tabular}{l|c|c|c|c|c|c|c|c|c}
\hlineB{2}
\textbf{Task} &\multicolumn{2}{c|}{\textbf{Clone Det.}} &\multicolumn{7}{c}{\textbf{Bug Detection}} \\\hline
\textbf{Dataset} &\textbf{P104} &\textbf{J250} &\multicolumn{3}{c|}{\textbf{RV}} &\multicolumn{2}{c|}{\textbf{D2A}} &\multicolumn{2}{c}{\textbf{CXG-DV}} \\
\hlineB{2}
\textbf{Metric} &\multicolumn{2}{c|}{\textbf{MAP@R}} &\textbf{P.} &\textbf{R.} &\textbf{F1} &\textbf{Acc} &\textbf{F1} &\textbf{Acc} &\textbf{F1} \\
\hlineB{2}
w/o LTSP & 91.3& 86.1& \textbf{48.7}& 46.9& 47.8& 60.0& 56.1& 63.4& 56.1\\\hlineB{2}
\rowcolor{rowgray}
w/ LTSP & \textbf{91.5}& \textbf{86.5}& 47.8& \textbf{49.3}& \textbf{48.6}& \textbf{62.1}& \textbf{67.1}& \textbf{63.7}& \textbf{58.3}\\
\hlineB{2}
\end{tabular}}
\end{table}

\noindent\textbf{Results.} \revision{In Table~\ref{tab: ltsp_ablation}, we conclude that removing LTSP hurts the model's performance in general, degrading clone retrieval performance slightly and F1 of bug detection significantly. The results empirically reveal the necessity of learning code structures for better code representations and the effectiveness of our proposed LTSP objective in learning such information.} 

\RS{3}{\revision{\tool's LTSP pre-training objective effectively improves the model's performance in downstream tasks by guiding the model to learn code structures.}}

\subsection{RQ4. Applying \toolbf to Existing Pre-trained Code Models}
\label{subsec:rq4_concord_variation}

As we introduced in \S\ref{subsubsec:learn_code_structure_and_functionality}, \tool's two-phase pre-training strategy leaves the flexibility of replacing \tool's first phase with other BERT-like pre-trained code models. In this RQ, we explore \tool's extensibility by applying it to existing syntax-based code models. We expect that \tool is able to improve the performance of pre-trained code models, and meanwhile, these existing models can help \tool to extend to more tasks.

\noindent\textbf{Setup.} We choose the two most popular models for experiments: CodeBERT and GraphCodeBERT. Specifically, we load the pre-trained weights from CodeBERT and GraphCodeBERT to initialize the Transformer-encoder layers within \tool architecture, and further train these models with \tool's multi-task second phase. We name these variations as CONCORD-CB (initialized with CodeBERT) and CONCORD-GCB (initialized with GraphCodeBERT). We fine-tune these models on the same downstream tasks discussed in \S~\ref{subsec:rq1}. 

Another benefit of using CodeBERT and GraphCodeBERT is that they are pre-trained with bi-modal datasets, where natural language and code both exist, so that we could extend \tool to bi-modal downstream tasks with these models.

\noindent\textbf{Extending \toolbf to Bi-modal Task: Code Search.} \label{subsec:code_summary_and_search_exp_setting}
Code search is the task of retrieving programs that match the natural language description, from tens of thousands of candidates. It requires the model's capacity of capturing semantic similarity of texts (between natural languages and code) rather than only syntactic similarity. Empowered by the bi-modal pre-trained models, we expect \tool to perform well in code search, as its semantic-aware contrastive learning aligns well with this task.

CodeSearchNet~\cite{husain2019codesearchnet} is the most popular benchmark for code search. It pairs each function with a natural language description, relating to the code comments. CodeSeachNet does not have datasets for C and C++, so we choose to evaluate \tool on its Java dataset. Again, we take the reconstructed benchmark from CodeXGLUE, which has 164,923 / 5,183 / 10,955 samples for train / valid / test splits respectively.
We follow the benchmark's design, using MRR (mean reciprocal rank) as the evaluation metric, and use the originally reported scores\footnote{Reference: \url{https://github.com/microsoft/CodeBERT/tree/master/GraphCodeBERT/codesearch}} for comparison.

\noindent\textbf{Results.}Table~\ref{tab: other_model_comp} summarizes the comparison between the existing pre-trained models and their \tool enhanced variants. We could see that \tool variants win on all tasks with a clear margin. These results empirically prove that \tool can significantly improve the performance of existing code models. Also, the code search result also reveals that \tool can be extended to multi-modal tasks and perform well by loading a multi-modal pre-trained model.

\begin{table}[!htp]\centering
\caption{\small Results of applying \toolbf to existing pre-trained code models.}\label{tab: other_model_comp}
\small
\resizebox{\linewidth}{!}
{
\begin{tabular}{l|c|c|c|c|c|c|c|c}
\hlineB{2}
\textbf{Task} &\multicolumn{2}{c|}{\textbf{Clone Det.}} &\multicolumn{5}{c|}{\textbf{Bug Detection}} &\textbf{Search} \\\hline
\textbf{Dataset} &\textbf{P104} &\textbf{J250} &\textbf{RV} &\multicolumn{2}{c|}{\textbf{D2A}} &\multicolumn{2}{c|}{\textbf{CXG-DV}} &\textbf{CSNet} \\\hlineB{2}
\textbf{Metric} &\multicolumn{2}{c|}{\textbf{MAP@R}} &\textbf{F1} &\textbf{Acc} &\textbf{F1} &\textbf{Acc} &\textbf{F1} &\textbf{MRR} \\\hlineB{2}
CodeBERT &82.7 &81.1 &47.3 &59.2 &63.6 &63.4 &53.1 &67.6 \\
\rowcolor{rowgray}
CONCORD-CB &\textbf{89.3} &\textbf{85.1} &\textbf{48.7} &\textbf{61.5} &\textbf{65.5} &\textbf{64.6} &\textbf{60.6} &\textbf{69.7} \\\hlineB{2}
GraphCodeBERT &86.7 &84.3 &46.6 &61.0 &66.1 &62.9 &56.3 &69.1 \\
\rowcolor{rowgray}
CONCORD-GCB &\textbf{91.6} &\textbf{84.8} &\textbf{47.2} &\textbf{62.3} &\textbf{70.0} &\textbf{64.2} &\textbf{60.1} &\textbf{70.5} \\
\hlineB{2}
\end{tabular}}
\end{table}

\RS{4}{\tool framework is \textbf{flexible }to be adapted with existing pre-trained code models and \textbf{improve their performance} in downstream tasks.}

\subsection{RQ5. Semantic-aware Code Representations}
\label{subsec:rq5_better_rep}

\begin{figure}[!ht]
    \centering
    \begin{subfigure}[t]{0.49\columnwidth}
     \includegraphics[width=\columnwidth]{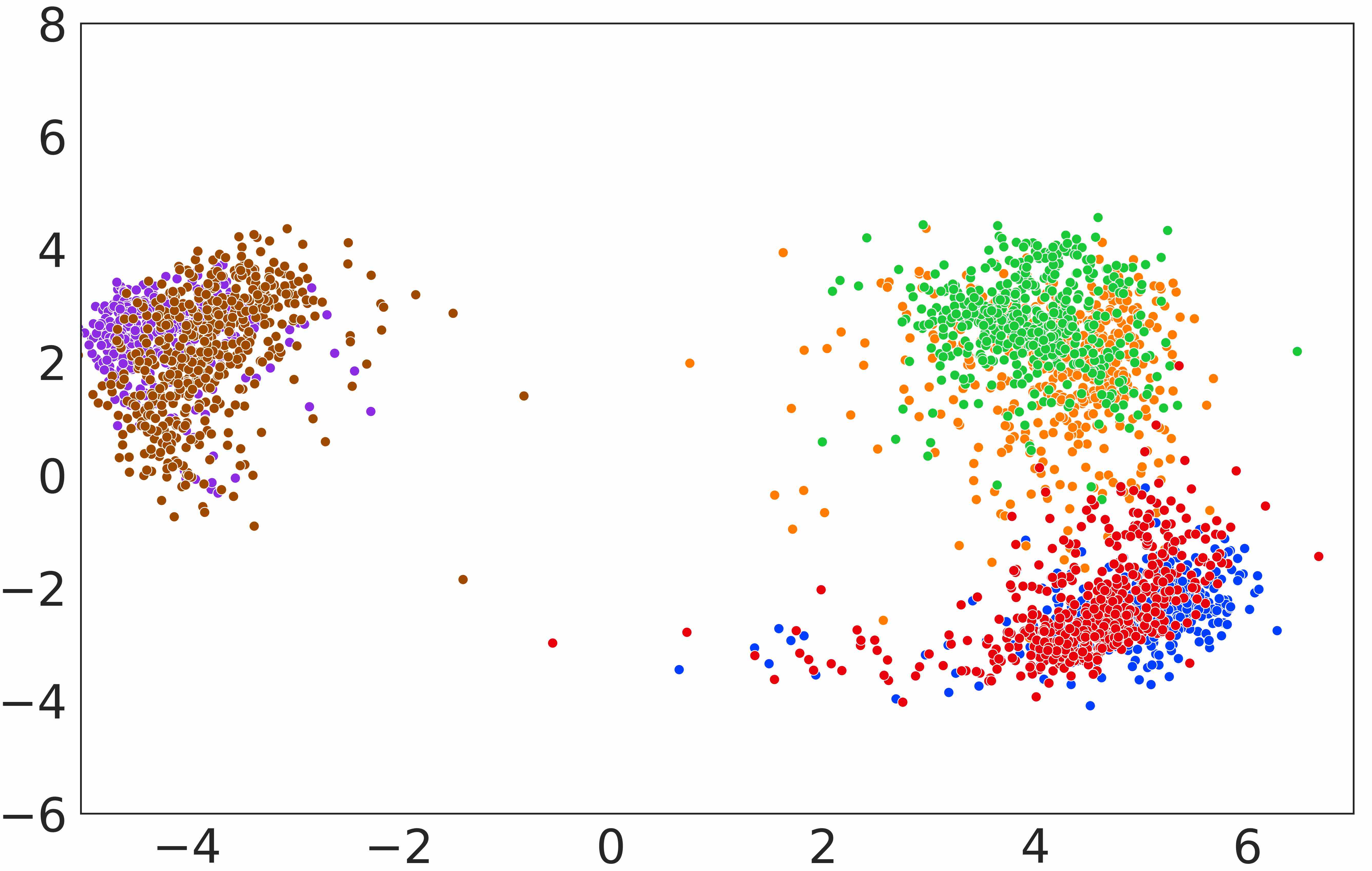}
     \caption{\small Existing Code Model}
     \label{fig:codebert_pca}
    \end{subfigure}
    \begin{subfigure}[t]{0.49\columnwidth}
     \includegraphics[width=\columnwidth]{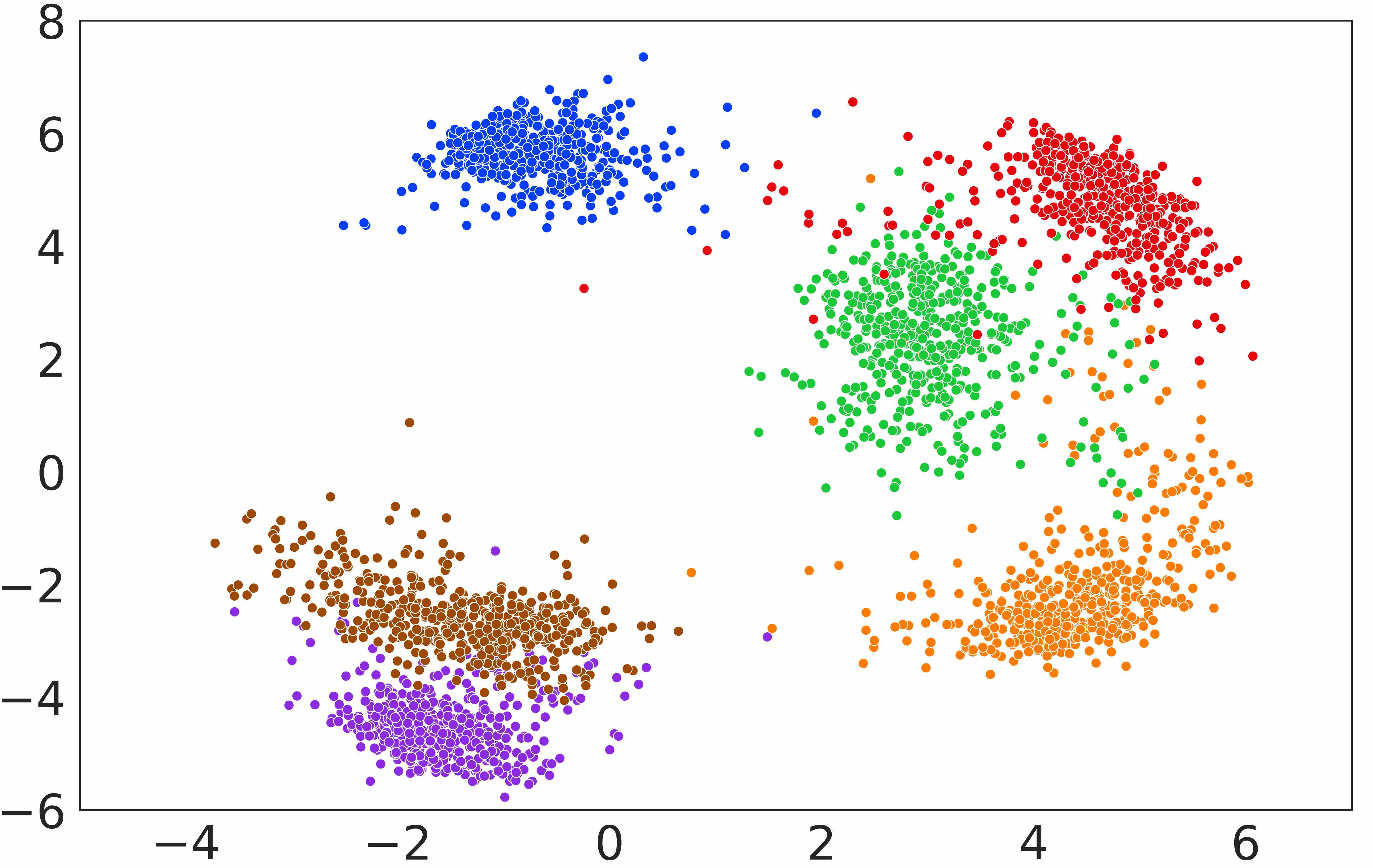}
     \caption{\small CONCORD}
     \label{fig:concord_pca}
    \end{subfigure}
    \caption{\small Visualization of data points in clone detection using PCA. Different colors represent distinct coding challenges in POJ-104}
    \label{fig:vis_clone}
\end{figure}

When we further study and visualize the code representations, we found that, even after fine-tuning, existing code models still struggle to encode programs based on semantic similarity. We take six coding challenges from the test split of POJ-104 dataset, and generate these samples' representations using the models \emph{fine-tuned} on the POJ-104 to study their distributions. For better comparison and visualization, we use principle component analysis (PCA)~\cite{hotelling1936pca} to reduce representations' dimensions and plot the 2-d data points in Figure~\ref{fig:vis_clone}: \ref{fig:codebert_pca} is generated by CodeBERT, and \ref{fig:concord_pca} is generated by CONCORD-CodeBERT, and each color represents one coding challenge. Clearly, CodeBERT's representations of distinct coding challenges significant overlap, which makes it difficult for the model to retrieve the semantic clones and tend to make mistakes. In contrast, CONCORD-CodeBERT's representations have clear boundaries among different clusters, so retrieving clones of the same code challenging becomes efficient. 

Motivated by this finding, we study the general quality of code representations of existing pre-trained code models and their enhanced variant with \tool. We design experiments to compare the similarity of code representations between the original code and its clone / clone-deviant / random code. Ideally, the model with decent awareness of code semantics should generate code representations following the order of \textsc{Sim[Original, Clone]} $>$ \textsc{Sim[Original, Deviant]} $>$ \textsc{Sim[Original, RandomCode]}, where \textsc{Sim[]} is the cosine similarity of two vectors, since the clone is semantically equivalent to the original code, and clone-deviant is buggy but textually more similar than other random code in the wild.

\noindent\textbf{Setup.} We randomly sampled 10,000 samples from the held-out dataset to ensure the programs are from real developers, and \tool models have never seen these code during training. Then we run \tool's data augmentation tool to pair each sample, $x$, with a clone, $x^+$, and a deviant, $x^-$. We end up with 10k triplets of ($x$, $x^+$, $x^-$) as the dataset of this section. Then we encode every program in the dataset with different pre-trained models we are studying. For the convenience of discussion, we define the set of original programs as $X = \{x_i | x_i ~\text{is original code}\}$, where $i \in [0,10,000)$, and augmented programs as $\hat{X} = \{\hat{x}_j | \hat{x}_j ~\text{is clone or deviant}\}$, where $j \in [0, 20,000)$, since one original program is augmented by two counterparts (clone and deviants). Finally, we exhaustively compute the cosine similarity of every possible pair of $\{x_i, \hat{x}_j\}$. All the experiments are conducted with the \emph{zero-shot} setting.

\noindent\textbf{Metrics.} We use two metrics to evaluate this RQ. First, we compute average pair-wise similarity to measure the similarity of original code and its clone, deviant and other random code within the same dataset. Specifically, we compute 
\begin{align*}
    \text{Avg. Clone Pair Similarity} &= \mathbb{E}_{x \in X}SIM(x, x^+) \\
    \text{Avg. Deviant Pair Similarity} &= \mathbb{E}_{x \in X}SIM(x, x^-)\\
    \text{Avg. Random Pair Similarity} &= \mathbb{E}_{x \in X}\mathbb{E}_{\hat{x} \in \hat{X}/\{x^+, x^-\}}SIM(x, \hat{x})
\end{align*}
Second, we evaluate the models' capacity of retrieving semantically similar code using the learned code representations. For each $x \in X$, we retrieve its Top-1 similar code from $\hat{X}$, and we check how often the retrieved code is $x$'s clone (\ie $x^+$), deviant (\ie $x^-$), and irrelevant code respectively.

\begin{table}[!htp]\centering
\caption{\small Code representation similarities between the original code and clone / clone-deviant / irrelevant code. Pair-wise Similarity is compared horizontally ($\leftrightarrow$) within the same model; Top-1 similar is compared vertically ($\updownarrow$) across models.}\label{tab: better_rep}
\small
\resizebox{\linewidth}{!}
{
\begin{tabular}{l|ccc|c|c|c}\hlineB{2}
\multirow{2}{*}{\textbf{Model}} &\multicolumn{3}{c|}{\textbf{Avg. Pair-wise Similarity ($\leftrightarrow$)}} &\multicolumn{3}{c}{\textbf{Top-1 Similar @ 20K ($\updownarrow$)}} \\\cline{2-7}
&\textbf{Clone} &\textbf{Deviant} &\textbf{Rand.} &\textbf{Clone} &\textbf{Deviant} &\textbf{Rand.} \\\hlineB{2}
CodeBERT &99.3 &\textbf{99.8} &97.5 &12.1 &85.8 &2.1 \\\cline{2-4}
CONCORD-CB &\textbf{95.5} &25.8 &2.9 &\textbf{97.1} &\textbf{1.0} &\textbf{1.9} \\\hlineB{2}
GraphCodeBERT &90.5 &\textbf{97.2} &72.9 &9.2 &87.4 &3.4 \\\cline{2-4}

CONCORD-GCB &\textbf{95.7} &22.7 &2.2 &\textbf{97.6} &\textbf{0.8} &\textbf{1.6} \\
\hlineB{2}
\end{tabular}
}
\end{table}

\noindent\textbf{Results.} Results are in Table~\ref{tab: better_rep}. Unfortunately, both CodeBERT and GraphCodeBERT assign higher similarity score to (original, deviant) pairs than (original, clone) pairs, and purely MLM-based CodeBERT assigns very high scores even to (original, random) pairs. These results show that existing code models have a weak sense of semantic similarity of source code, as they always encode syntactically similar code as closer representations. On the contrary, after training these models with \tool's approach, their representations become more aware of semantic similarity: both CONCORD-CodeBERT and CONCORD-GraphCodeBERT regard (original, deviant) pairs as less similar than (original, clone) pairs, and give (original, random) pairs the least similarity scores. Similarly, syntax-based models tend to wrongly retrieve the deviant as the top-1 similar program in most cases, while \tool can always pinpoint the semantically equivalent programs from tens of thousands of candidates.

\RS{5}{\tool effectively improves syntax-based models to learn better code representations for identifying semantic similarity.}

\section{Related Work}
\label{sec2:related}
\paragraph{\textbf{Self-supervised Pre-training for Code}}
Researchers have been passionate about pre-training Transformer models for source code. There are three main architectures for existing models: Encoder-only~\cite{feng2020codebert, guo2021graphcodebert, xin2021syncobert,bui2021corder,cubert,buratti2020cbert, ding2021disco}, Decoder-only~\cite{chen2021codex,xu2022systematic, austin2021synthesis}, and Encoder-decoder~\cite{niu2022sptcode,ahmad2021plbart,guo2022unixcoder,Li2022CompetitionLevelCG, chakraborty2021natgen}. Encoder-only models are commonly pre-trained with cloze tasks (\eg masked language model) and sequence understanding tasks (\eg next statement prediction). Decoder-only models are mostly trained with autoregressive, left-to-right language model (LM)
Encoder-Decoder models are pre-trained with different tasks including denoising autoencoding to reconstruct the wrongly permuted tokens~\cite{ahmad2021plbart}, predicting missing identifiers~\cite{wang2021codet5}, recovering method names~\cite{niu2022sptcode}, etc.
In recent years, with the rapid development of computing 
devices, such as GPUs and TPUs, researcher also shed light on the incredible power of extremely large Transformer models (up to hundreds of billions of parameters) for understanding and generating code~\cite{austin2021synthesis, chen2021codex, github-2021-copilot, Li2022CompetitionLevelCG}.

\paragraph{\textbf{Contrastive Learning for Code.}}
Most recently, self-supervised contrastive learning has gained a lot of interest in learning source code representations~\citep{chen2021varclr, bui2021corder, xin2021syncobert, lu2022reacc, ding2021disco, guo2022unixcoder, jain2020contrastive}. \revision{Contrastive learning models for source code typically include two steps: (1) augmenting datasets with semantically equivalent programs as positive samples and contradictory programs as negative samples. (2) learning to maximize the vector similarity of equivalent samples and minimize the similarity of contradictory samples.
Ding \etal proposes DISCO~\cite{ding2021disco} that generates the functionally equivalent code
 by renaming identifiers and permuting independent statements, and involves small security bugs as hard-negative samples. Besides the contrastive learning objective, DISCO also introduces NT-MLM to capture the code syntax. Corder~\cite{bui2021corder} designs semantically preserving AST transformations to produce positive samples. ContraCode~\cite{jain2020contrastive} uses the compiler to conduct the source-to-source compilation, \emph{a.k.a.,}transpilation, which is originally for code optimization and obfuscation, to generate positive counterparts for JavaScript. We will compare \tool with these state-of-the-art contrastive learning code models in \S~\ref{sec:discussion}.

}

\revision{
\section{Discussion}
\label{sec:discussion}

In this section, we discuss and compare \tool with several most relevant code models using contrastive learning: DISCO~\cite{ding2021disco}, Corder~\cite{bui2021corder}, and ContraCode~\cite{jain2020contrastive}. The comparison will be explained with respect to (1) data augmentation and (2) pre-training strategy.

\noindent\textbf{DISCO.} For data augmentation, DISCO's oversimplified data augmentation approach hurts its performance. Its positive heuristics for generating functionally equivalent code focus on renaming variables and functions with abstract names like "VAR\_0" and "FUNC\_0", which are rare in real programs, resulting in unnatural clones. In contrast, \tool incorporates real developers' cloning patterns into pre-training through several clone generation rules, inspired by patterns drawn from the existing studies~\cite{chakraborty2021natgen, gopstein2017understanding, gopstein2018prevalence, gopstein2020thinking, roy2009nearmiss, uddin2011nearmiss} and pre-defined clone types (Type- 1,2,3,4). As a result, Table~\ref{tab:clone_detection} and Table~\ref{tab: bug_detection} show that, in the same evaluation setup, \tool outperforms DISCO with a clear margin.

Beyond downstream tasks, we also compare the impacts of DISCO's and CONCORD's data augmentation on code representations quality, regarding identifying semantic similarity. We augment the original dataset with DISCO data augmentation, and train \tool on top of it (called CCD-DISCO). We reuse the zero-shot setup discussed in \S~\ref{subsec:rq5_better_rep} that first encodes the program with pre-trained models, and then computes the average of \textsc{Sim[Original, Clone]} and retrieves the top-1 similar code. Table~\ref{tab: better_rep_disco_data_aug} shows that, due to the oversimplified and unnatural transformations used to generate clones, CCD-DISCO has significantly lower cosine similarity between the representation of the original code and the clone, and a higher chance of failure in retrieving clones with the zero-shot setting when compared to \tool.

\begin{table}[!htp]\centering
\caption{\small \revision{Comparison of the quality of code representations in identifying code similarity between CCD-DISCO and \toolbf. $\uparrow$ indicates that a larger value represents a better representation, while $\downarrow$ indicates that a smaller value represents a better representation.}}\label{tab: better_rep_disco_data_aug}
\small
\resizebox{\linewidth}{!}
{
\begin{tabular}{l|c|c|c|c}\hlineB{2}
\multirow{2}{*}{\textbf{Model}} & \textbf{Avg. Similarity}  &\multicolumn{3}{c}{\textbf{Top-1 Similar @ 20K}} \\\cline{3-5}
&\textbf{ with Clone ($\uparrow$)} &\textbf{Clone ($\uparrow$)} &\textbf{Deviant ($\downarrow$)} &\textbf{Rand. ($\downarrow$)} \\\hlineB{2}
CCD-DISCO &71.7 &64.4 &3.7 &31.8 \\\cline{1-5}
CONCORD &\textbf{94.8} &\textbf{95.7} &\textbf{2.2} &\textbf{2.1} \\\hlineB{2}
\end{tabular}
}
\end{table}

DISCO's pre-training strategy also has limitations compared to \tool. First, it couples the multi-task pre-training into one single phase. This makes it less flexible to leverage large pre-trained code models, and also empirically less effective as the randomly initialized model struggles to learn multiple perspectives of source code at once. As a comparison, \tool proposes multi-phase pre-training that first learns the general perspective of code text and then specializes to learn the syntax and semantics during the second phase. Second, DISCO pre-trains a mono-lingual model for each programming language (PL) that fails to unify the common knowledge across distinct PLs and degrades the quality of the learned code representations~\cite{athiwaratkun2023multilingual}. Training mono-lingual models separately also wastes training resources. In contrast, \tool is a multi-lingual model that learns comprehensive code representations across distinct PLs with cheaper training costs. Third, \tool's LTSP is by design more capable of encoding code structures than DISCO's NT-MLM, as the former predicts local ASTs for the whole program while the latter only reconstructs the node type for the masked token.

\noindent\textbf{Corder.} For data augmentation, Corder proposes AST transformations to synthesize semantic preserving samples. However, Corder does not consider any types of hard negative samples that are guaranteed to behave differently from the original code. This overlook might weaken the model's capacity in contrasting the semantic similarity of code, especially in differentiating the benign and buggy samples that are textually similar (\eg Figure~\ref{fig:original_code} \emph{vs.} Figure~\ref{fig:clone_deviant}). In contrast, \tool enhances the learning with clone-deviants as hard negative samples, which include bugs that maliciously change the original program behaviors. During pre-training,
Corder takes AST-based intermediate representation as input and ignores learning the source code text directly. Such ignorance could make the model less capable of understanding the rich semantics underneath the source code text, such as variable/function names and comments, which are the main resources to expose developers' intentions during coding~\cite{casalnuovo2020does, casalnuovo2020programmers, casalnuovo2020theory}, and consequently, degrade the quality of learned code representations.

\noindent\textbf{ContraCode.} ContraCode smartly leverages the off-the-shelf compiler to generate optimized or obfuscated programs as semantic preserving samples. While the optimized and obfuscated code provides precise and formal semantics~\cite{casalnuovo2020theory}, they tend to be unnatural, introducing data structures and variable names that are not commonly used in human-written programs. Existing studies have argued that such formal but unnatural programs are less favorable to human developers~\cite{casalnuovo2020does, casalnuovo2020programmers} and obstruct the code models' learning~\cite{chakraborty2021natgen}. Also, ContraCode does not generate semantically contradicting programs as hard negative samples. In contrast, \tool imitates the developers' cloning patterns to augment the dataset with clones and clone-deviants, better aligning with human-written programs. For pre-training, ContraCode does not learn code syntax, such as ASTs, which might hurts the model's performance. ContraCode performs poorly in our evaluation datasets (\eg its best variant only reports 65.6 MAP@R on POJ-104) due to the above limitations as well as other practical restrictions, such as the model size being too small to compete with other baselines we discussed.

}

\vspace{-1mm}
\section{Threats to Validity}
\label{sec:threats}

We argued that incorporating real developers' coding patterns into pre-training helps to learn better, generic code representations. As a proof of concept, we choose clone-related coding behaviors as our main focus, since code clones are happening all the time in daily development. However, there are still other interesting patterns, such as how developers name variables/functions~\cite{chen2021varclr, Ahmed2021MultilingualTF}, that will help deep-learning models understand source code and can be integrated into the pre-training. Also, our data augmentation only generate the variants of the same programming language as the original code, and our model might have limited capacity in detecting cross-lingual semantic clones.

Our data augmentation is trying to imitate the developers' cloning behaviors and have designed multiple transformation rules, but they may not cover all the clone patterns, as programming is a very personal activity~\cite{knuth1984literate}, and different persons can implement the same function with drastically distinct algorithms. 
It is also very difficult to guarantee that clone-deviants will exhibit malicious behaviors, but we expect the model to be sensitive to unexpected changes as they are highly likely to introduce software flaws. In addition, during contrastive learning, we consider in-batch samples as random negative samples, since we assume that samples inside the same batch do not have similar functionalities. This assumption might not hold for 100\% cases. However, because the batch is randomly sampled from millions of programs, it should be very rare that the in-batch programs share the same functionalities.

Another limitation is that \tool is pre-trained with multiple objectives, such that weights of different loss functions might have impacts on the quality of pre-training~\cite{gao2021simcse}. We design the weights to be $\lambda_1=1.0$, $\lambda_2=0.1$, $\lambda_3=1.0$ based on the intuition that giving high weights to LTSP might force the model to focus too much on code syntax and diminish the impacts of semantic-aware contrastive learning. \revision{We also conducted experiments with $\lambda_2=0.1$, $0.5$, and $1.0$, and $\lambda_2=0.1$ reports the best performance on downstream tasks but their difference is not significant: \eg $\lambda_2=0.1$ reports 91.5/86.5 MAP@R in the clone detection datasets, while $\lambda_2=0.5$ reports 91.5/86.4 MAP@R, and $\lambda_2=1.0$ reports 91.3/86.3 MAP@R, respectively. Pre-training code models is expensive, so we did not exhaustively search for the best weights for each objective.} Other weights might improve \tool's performance.

\vspace{-2mm}
\section{Conclusion}
\label{sec:conclusion}

In this paper, we incorporate into pre-training a common developer practice, copy/paste, to improve the quality and efficiency of learning code representations. We first introduce an automated tool to augment the code datasets with both semantic clones and buggy clone-deviants. With these augmented datasets, we pre-train \tool with MLM, LTSP, and CLR objectives. Our evaluation reveals the effectiveness and efficiency of our approach, showing that even with much cheaper training expenses, \tool still outperforms SOTA code models. In addition, \tool's approach can easily be applied to existing code models to improve their code representation quality.

\vspace{-2mm}
\section*{Acknowledgments}
We appreciate all the anonymous reviewers for their thoughtful feedback and suggestions to improve this work.

This work was supported in part by an IBM Ph.D. Fellowship, an IBM Faculty Award, NSF grants CCF-1815494, CCF-210740, CCF-1845893, IIS-2221943, and DARPA/NIWC Pacific N66001-21-C-4018. Any opinions, findings,
conclusions or recommendations expressed herein are those
of the authors and do not necessarily reflect those of the US
Government, NSF, DARPA, or IBM.

\balance
\bibliographystyle{ACM-Reference-Format}
\bibliography{main.bib}

\end{document}